\documentclass{emulateapj}

\slugcomment{Submitted to the Astrophysical Journal}

\shorttitle{Intrinsic electromagnetic variability in celestial objects containing rapidly spinning black holes}
\shortauthors{Fan Zhang}

\usepackage{amsmath,amssymb}
\usepackage[normalem]{ulem}
\usepackage{bm}
\usepackage{comment}
\usepackage{xcolor}
\usepackage{soul}
\usepackage{hyperref}
\usepackage[percent]{overpic}
\usepackage{overpic}
\usepackage{natbib}
\usepackage[export]{adjustbox}

\begin{document}

\newcommand{\beq}{\begin{equation}}
\newcommand{\eeq}{\end{equation}}
\newcommand{\bea}{\begin{eqnarray}}
\newcommand{\eea}{\end{eqnarray}}
\newcommand{\ba}{\begain{align}}
\newcommand{\ea}{\end{align}}
\newcommand{\ph}{\phantom}
\newcommand{\bma}{\begin{pmatrix}}
\newcommand{\ema}{\end{pmatrix}}
\newcommand{\E}{\mathcal{E}}
\newcommand{\B}{\mathcal{B}}

\renewcommand{\baselinestretch}{1.1}
\newcommand{\R}[1]{\textcolor{red}{#1}}
\newcommand{\fan}[1]{\textcolor{blue}{#1}}

\title{Intrinsic electromagnetic variability in celestial objects \\ containing rapidly spinning black holes}
\author{Fan Zhang  \altaffilmark{1,2}}

\altaffiltext{1}{Gravitational Wave and Cosmology Laboratory, Department of Astronomy, Beijing Normal University, Beijing 100875, China}
\altaffiltext{2}{Department of Physics and Astronomy, West Virginia University, PO Box 6315, Morgantown, WV 26506, USA}

\date{\today}

\begin{abstract}
Analytical studies have raised the concern that a mysterious expulsion of magnetic field lines by a rapidly-spinning black hole (dubbed the black hole Meissner effect) would shut down the Blandford-Znajek process and quench the jets of active galactic nuclei and microquasars. This effect is however not seen observationally or in numerical simulations. Previous attempts at reconciling the predictions with observations have proposed several mechanisms to evade the Meissner effect.  In this paper, we identify a new evasion mechanism and discuss its observational significance. Specifically, we show that the breakdown of stationarity is sufficient to remove the expulsion of the magnetic field at all multipole orders, and that
the associated temporal variation is likely turbulent due to the existence of efficient mechanisms for sharing energy across different modes. Such an intrinsic (as opposed to being driven externally by, e.g., changes in the accretion rate) variability of the electromagnetic field can produce the recorded linear correlation between microvariability amplitudes and mean fluxes, help create magnetic randomness and seed sheared magnetic loops in jets, and lead to a better theoretical fit to the X-ray microvariability power spectral density.
\end{abstract}


\maketitle

\section{Introduction} \label{sec:Intro}
Building on the work of \cite{Goldreich:1969sb}, \cite{1977MNRAS.179..433B} (BZ) proposed a mechanism that extracts the rotational energy of spinning black holes through a Penrose-like process, which has since become one of the most widely accepted contender for explaining the energy source of out-flowing jets, seen in active galaxy nuclei (AGN) containing supermassive black holes and microquasars containing stellar mass holes. For example, the measured jet power's dependence on the black hole spin is consistent with BZ predictions (\cite{2012MNRAS.419L..69N,2013ApJ...762..104S}). In order to operate effectively, the BZ mechanism requires the near-horizon region to be sufficiently and appropriately magnetized. From the Penrose process point of view, the magnetic field lines should thread through the infinite redshift surface enclosing the ergosphere, while with the membrane paradigm description (\cite{Thorne-Price-MacDonald}) of the BZ process, one would need them to thread through the event horizon. 

\begin{figure}[t,b]
  \centering
\begin{overpic}[width=0.472182\columnwidth]  {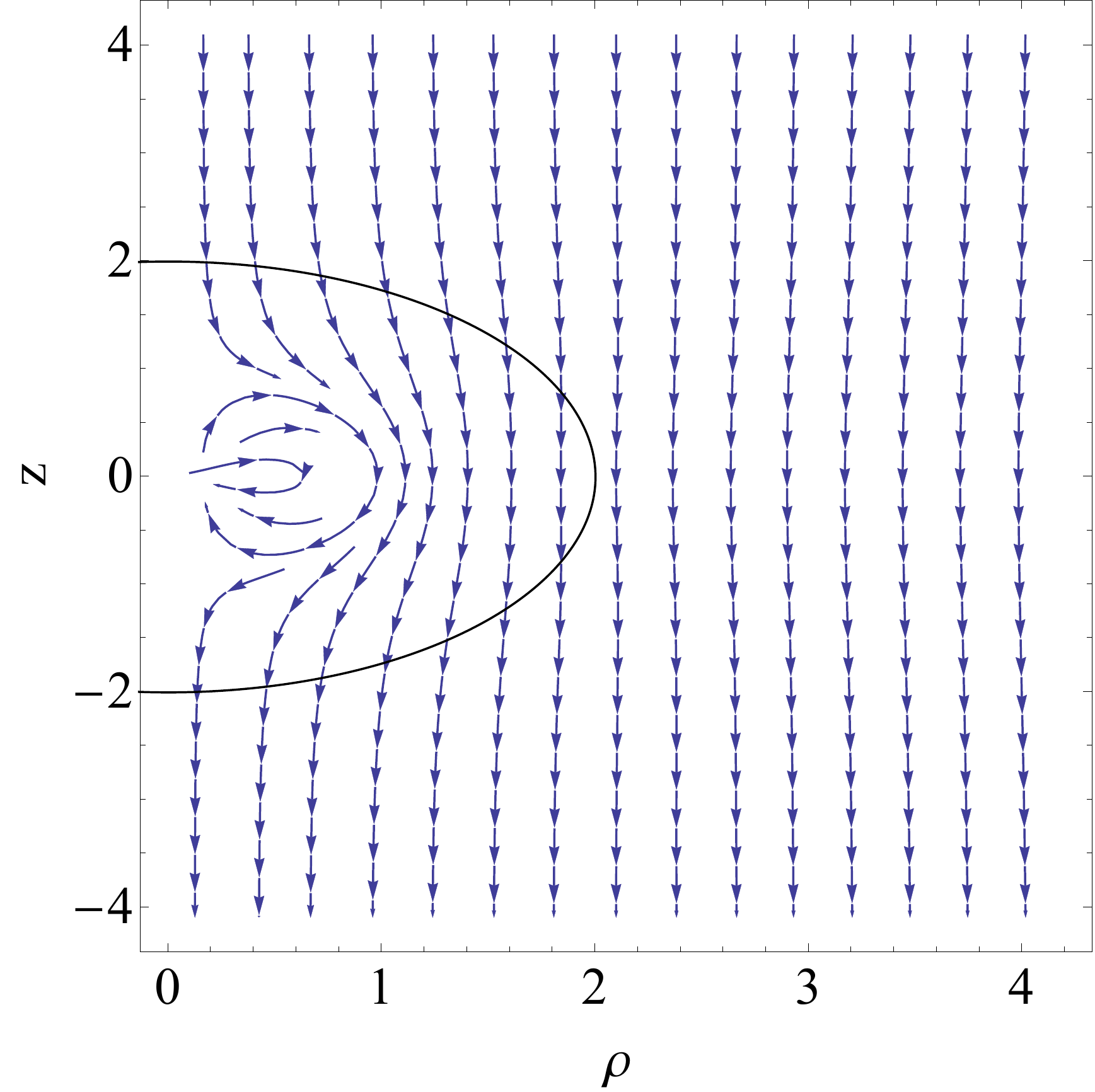}
\put(90,0){(a)}
\end{overpic}
\begin{overpic}[width=0.49\columnwidth]  {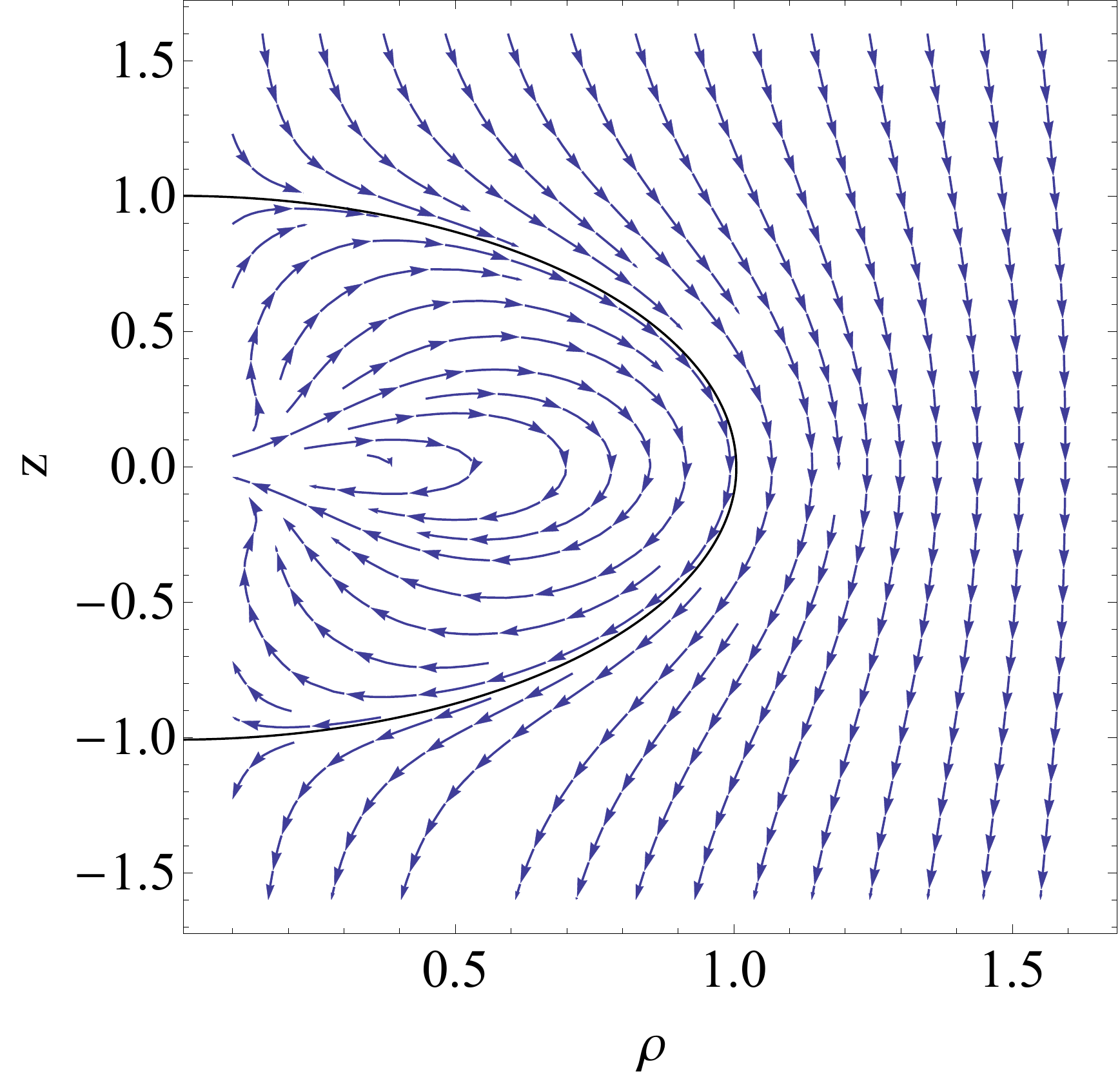}
\put(90,0){(b)}
\end{overpic}
\begin{overpic}[width=0.49\columnwidth]  {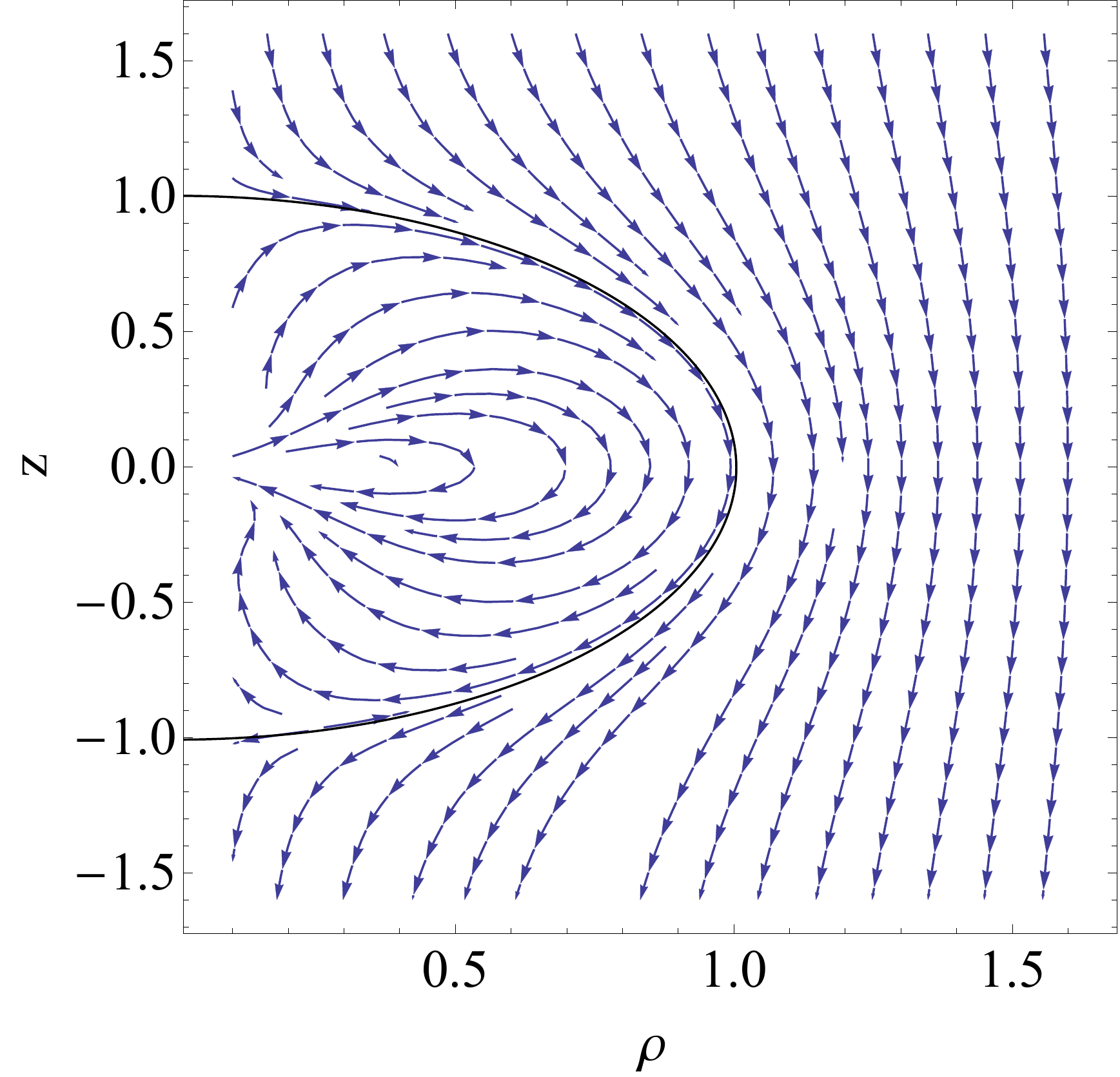}
\put(90,0){(c)}
\end{overpic}
\begin{overpic}[width=0.472182\columnwidth]  {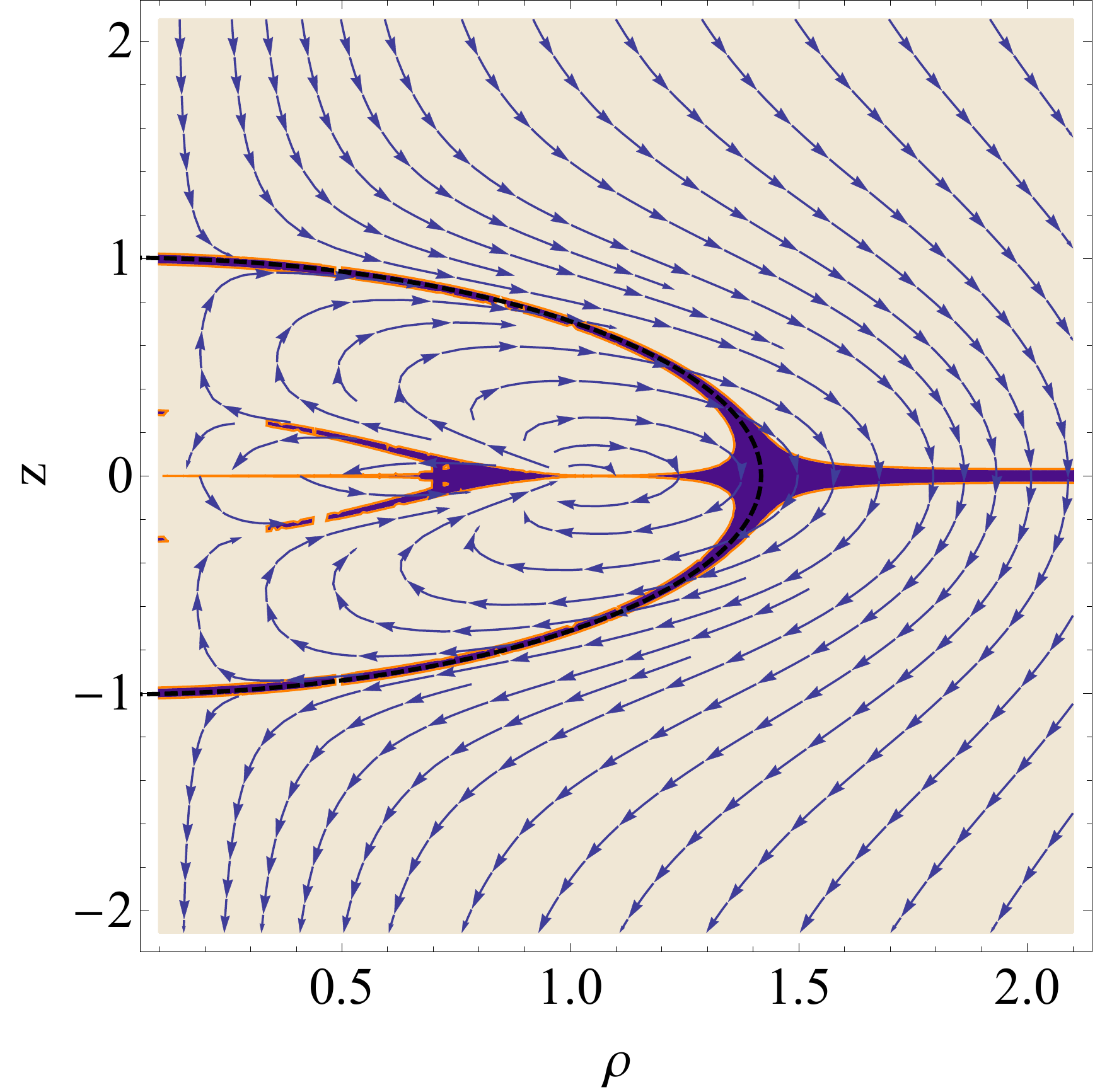}
\put(90,0){(d)}
\end{overpic}
  \caption{
  (a): The Wald solution in the cylindrical version of the Boyer-Lindquist coordinates for spin $a=0$. The event horizon is represented by the black semi-circle.
  (b): The Wald solution in the Boyer-Lindquist coordinates for spin $a=0.99999$. 
  (c): The Wald solution in the cylindrical version of the Ingoing Kerr coordinates for spin $a=0.99999$, the coordinate transformations from Boyer-Lindquist do not affect the $(\rho,z)$ plane projection of the field lines.
  (d): The Wald solution in the Kerr-Schild coordinates for spin $a=0.99999$. The event horizon is still represented by a semi-circle, but we have also shown an orange contour of the projection of the magnetic field onto the normal to the event horizon. The contour pieces straddle zero (the darker region corresponds to smaller projection value), so the flux density across the horizon actually vanishes, even though graphically it appears that the magnetic field lines thread through the horizon at places.
}
	\label{fig:WaldStream}
\end{figure}

Astrophysically, the magnetic field surrounding black holes is expected to be a largely poloidal one confined by external matter (\cite{1984RvMP...56..255B}), which can be approximated as being asymptotically uniform when studying the vicinity of the black holes. The simplest and frequently invoked prototype solution for such a configuration is the Wald solution (\cite{PhysRevD.10.1680}) to the vacuum (without charge or current) Maxwell equations. When it is observed that as the black hole spin approaches extremality, the magnetic field lines of this solutions are ``expelled" from the horizon (see \cite{King1975,Bicak:2006hs} and Fig.~\ref{fig:WaldStream}), instead of penetrating it as in the case of slow spin, concerns naturally arise that the fast spinning black holes won't be adequately magnetized so the BZ process shuts down, and the jets are quenched. 

\begin{figure}[t,b]
  \centering
\begin{overpic}[width=0.85\columnwidth]{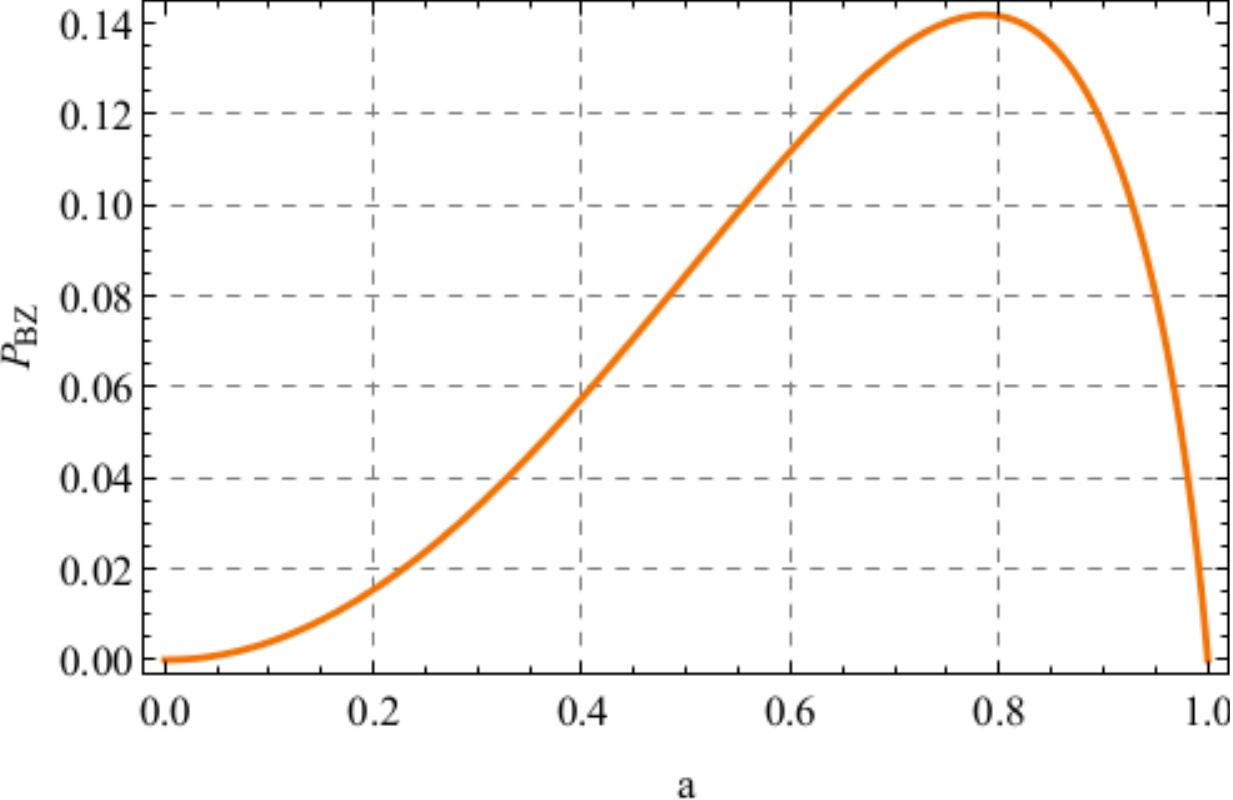}
\end{overpic}
  \caption{The power extracted from the black hole through the BZ process as predicted with Eq.~\eqref{eq:BZP}, setting black hole mass $M=1$, and the asymptotic magnetic field strength $B=1$. 
}
	\label{fig:BZPower}
\end{figure}

Although the complete expulsion of the field lines is only at extremality, the effect does bleed into the entire high spin regime, causing a severe suppression of the jet power $P_{\text BZ}$ from the BZ process, which is proportional to the quadrature of the magnetic flux across the horizon. Substituting in the Wald solution, we obtain $P_{\text BZ}$ as a function of the dimensionless spin $a$ as (\cite{Bicak85}) 
\bea \label{eq:BZP}
P_{\text BZ} &=& \frac{\pi B^2 a^2 r_+^2}{32 M^2}  \left(1-\frac{a^4}{r_+^4}\right)^2\,, \notag \\ 
\text{where} \quad r_+ &=&M+\sqrt{M^2-a^2}\,,
\eea
which is plotted in Fig.~\ref{fig:BZPower}. The $a$ appearing in the bracket is from the magnetic flux density contribution, and we see that power suppression sets in after $a\sim 0.8$, which is well within the possible astrophysical range. Theoretical computations by \cite{Thorne:1974} show that thin accretion disks can spin a black hole up to $a\sim 0.998$ (although the numerical experiment by \cite{GammieEtAl:2004} shows that lower equilibrium values at around $0.9$ for particular magnetohydrodynamic (MHD) models are also possible). Within the BZ process, a near-extremal spin is also assumed when discussing the collimation of extracted energy into jets by an ion-supported torus (\cite{1982Natur.295...17R,1984RvMP...56..255B}). Observationally, high spins have been reported for many systems, using a variety of measurement techniques.  For example, the Fe K line profile implies $a\approx 1$ for 
XTE J1650–500 (\cite{2002ApJ...570L..69M,2004MNRAS.351..466M}), 
while \cite{McClintock:2006xd}, \cite{Gou:2011nq} and \cite{McClintock:2013vwa} have measured spins in excess of $0.95$ for the primaries in black hole X-ray binaries Cygnus X-1 and GRS 1915+105, through a continuum fitting method. Furthermore, when it comes to spin distribution among large populations, high spins are not rare. 
Summarizing  X-ray reflection spectroscopy results, \cite{Reynolds:2013rva} concluded that a significant number of supermassive black holes have $a>0.9$.
With a radiation efficiency study, \cite{2006ApJ...642L.111W} estimated that the spins of essentially all black holes in quasars within the redshift range of $0.4<z<2.1$ to be close to $1$.
Via a consideration of the quasar contribution to the X-ray background, \cite{2002ApJ...565L..75E} also arrived at the conclusion that ``most supermassive black holes must be rapidly rotating''. 

The expulsion of the magnetic field by rapidly-spinning black holes (named the black hole Meissner effect in analogy with superconductors) is therefore likely more than just a theoretical curiosity, and has to be examined carefully if we want to understand real celestial objects. One would of course suspect that this effect is an esoteric feature only of the Wald solution. However, subsequent studies by \cite{
Bicak76,
King77,
1980PhRvD..22.2933B,
Bicak85,
Bicak89,
1991JMP....32..714K,
1998PhRvD..58h4009C,
Karas00,
2007IAUS..238..139B,
2013CQGra..30l5008G,
2015arXiv151001911B,
Bicak2015} have greatly generalized the Wald solution to arbitrary axisymmetric solutions, to models containing the back-reaction of the electromagnetic field on spacetime, and to higher dimensions etc, and this Meissner effect persists. A precise characterization of this effect is given by \cite{Bicak85}, that black holes in extremality (in spin, charge, or a combination of both) expel external \emph{axisymmetric stationary} magnetic fields. One may also ask the question that if the energy extraction process occurs in the ergosphere, would that render the Meissner effect irrelevant? \cite{Penna:2014aza} argued that the magnetic horizon flux density is required to ensure a clear passage for the negative energy components in the Penrose-like BZ process to cross the horizon, and thus remain a necessity. \cite{PanYu2015} further showed that when perfectly-conducting plasma is present, the magnetic field lines crossing into the ergosphere must also intersect with the horizon, so the Meissner effect would in fact expel the field lines all the way out of the ergosphere. 

Despite the analytical predictions, numerical general relativistic MHD simulations see no sign of this effect (\cite{2004ApJ...611..977M,2013ApJ...765..113C,2012MNRAS.423.3083M,2013MNRAS.436.3741P,2012JPhCS.372a2040T}). Observationally, systems like microquasar GRS 1915+105 (spin $0.98$) do not exhibit signs of jet quenching, i.e. their jet power is consistent with a $a^2$ scaling, showing no additional $a$ dependent suppression in magnetic flux density (\cite{2012MNRAS.419L..69N,2013ApJ...762..104S}). Therefore, mechanisms for evading the Meissner effect are needed, and several are already proposed. While examining arbitrary stationary solutions, \cite{Bicak85} pointed out that solutions that are not axisymmetric do not experience the Meissner effect (briefly reviewed in Sec.~\ref{sec:Review} below). Later, \cite{Penna:2014aza} further identified the (split) monopole configuration as being able to evade it \footnote{As is noted in \cite{Penna:2014aza}, the infinite extremal Kerr throat argument utilized to elucidate the Meissner effect there is observer/coordinate dependent. We point out that \cite{Jacobson:2011ua} is a good reference that explains this issue. 
Also, the throat argument only requires the magnetic field lines to penetrate the horizon radially in the Boyer-Lindquist coordinates, saying nothing about their density's angular dependence. So while the monopole offers the simplest complying configuration, the results of \cite{Penna:2014aza} do not prohibit additional evasion mechanisms activating higher multipoles.}. 
\cite{2011MNRAS.412.2417T} also found that currents carried by plasma in magnetospheres enable higher multipole magnetic fields to penetrate the horizon in the case of extremally charged (but nonspinning) black holes. On the other hand, if one allows an extremally spinning black hole to be charged, an induced magnetic dipole field can thread through the horizon (\cite{1980PhRvD..22.2933B}). 

The remaining obvious loophole to exploit is the breaking of stationarity, which has escaped close scrutiny in previous studies concentrating on exact analytical solutions, due to the technical difficulty in handling fully dynamic electromagnetic fields in a curved spacetime. In Sec.~\ref{sec:TimeDep} below, we utilize semi-analytic arguments to show that intrinsic temporal variability (ITV) in the electromagnetic field can indeed serve as another route for the evasion of the Meissner effect. Furthermore, ITV may be a feature of prevalence among real astrophysical systems. 
From an observational perspective, we know that AGN and microquasar emissions are variable, so the electromagnetic field surrounding astrophysical black holes are not stationary. 
From an energetics point of view, ITV brings about a more effective extraction of the black hole rotational energy, some of which would become available to fuel further dynamic behaviour. 
So ITV may have a tendency to self-enhance (given the right conditions, see the end of Sec.~\ref{sec:Conclude} for more discussions), 
until an equilibrium with dissipation is reached (alternatively, if stable equilibrium is difficult to achieve, cycles of active and quiescent periods may alternate, causing ITV to appear intermittent). 
Together with the rotational energy, angular momentum of the black hole is also extracted, so ITV can help prevent the astrophysical black holes from over-spinning into naked singularities (or simply reaching extremality, which \cite{Aretakis:2012ei,Aretakis:2013dpa,Lucietti:2012sf} have suggested to be nonlinearly unstable), thus serving as an enforcer for cosmic censorship. In short, ITV may be an integral part of a self-consistent picture of astrophysical systems containing rapidly but sub-extremally spinning black holes. 

Assuming this, it is then interesting to examine what kind of more detailed associations can be made between ITV and observations. One feature of ITV is that when it enhances the horizon magnetic flux, the additional energy extracted would go equitably into many sibling modes of differing angular quantum numbers (see Sec.~\ref{sec:Turbulence}). Thus an efficient ``mode-coupling'' mechanism exists to promote the development of turbulence in the electromagnetic field, and subsequently ITV may be closely related to the stochastic microvariability of AGNs and microquasars. In Sec.~\ref{sec:Obs}, we discuss several potential applications to this observation: 
\begin{enumerate}
\item (Sec.~\ref{sec:VarFluxCorr}) ITV naturally produces a linear correlation between microvariability amplitudes and short term mean fluxes, which has been recorded in numerous studies. 

\item (Sec.~\ref{sec:Jet}) ITV possesses properties making it useful in modelling the magnetic field structure observed in jets by polarization studies. 

\item (Sec.~\ref{sec:PSD}) Using ITV and a photon leakage scheme, it is possible to obtain an improved theoretical fit to the X-ray microvariability power spectral density. \end{enumerate}
Finally, we conclude in Sec.~\ref{sec:Conclude}. The derivations below are in geometrized units with $c=1=G$, unless otherwise specified.

\section{The Meissner effect} \label{sec:Review}
We begin by reviewing the essential arguments of \cite{Bicak85} that led to the observation of the Meissner effect being generic for any arbitrary, axisymmetric, stationary field configuration with $l\geq 1$. We will then adapt them to the case of temporally variable configurations in the next section, and show that they also evade the Meissner effect, in much the same fashion as the rotationally asymmetric ones do. We will specialize to spin and not charge extremality, anticipating possible astrophysical applications. In addition, we will also restrict ourselves to vacuum (without plasma) solutions in the near-horizon regime, but allow arbitrary source charges and currents at further away, just as the original computation in \cite{Bicak85} assumed. This is mostly to ensure that the calculations remain tractable. On the other hand, the derivation of the BZ process included the presence of a so-called force-free plasma, in which the inertia of the charged particles are neglected because their stress-energy tensor is subdominant to that of the electromagnetic field. In reality, the particles do have mass and will not be able to travel at the speed of light and stay on the event horizon (a null surface by definition), so the particles sufficiently close to the horizon are on their way in, plausibly generating largely radial currents that do not greatly alter the radial component of the magnetic field and subsequently the horizon flux density, as compared to those in a very thin vacuum layer sourced by the same current and charge distribution outside (as the vacuum solutions examined below only need to satisfy horizon boundary conditions and we specialize to their limits when approaching the horizon, the vacuum layer can be very thin). 

The arguments of \cite{Bicak85} (and \cite{Bicak76} that it relies on) are based on the Newman-Penrose formalism. The underlying spacetime metric is that of Kerr 
\bea
ds^2 &=& -\frac{\Delta}{\Sigma}(dt-a \sin^2\theta d\phi)^2 
+\frac{\Sigma}{\Delta}dr^2 
+ \Sigma d\theta^2\, \notag \\
&&+ \frac{\sin^2\theta}{\Sigma} \left( (r^2+a^2)d\phi-adt\right)^2  \,,\notag \\
\Delta &=& r^2-2Mr+a^2=(r-r_+)(r-r_-)\,, \notag \\ \Sigma &=& r^2 + a^2 \cos^2\theta \,,
\eea
where $\{t,r, \theta,\phi\}$ are the Boyer-Lindquist coordinates, $a$ is the dimensionless spin parameter, $M$ is the black hole mass, and $r_+$ is the radius of the event horizon. A particularly nice Newman-Penrose null tetrad (basis to decompose vectors and tensors into their components with) exists in this spacetime, namely the Kinnersley tetrad (\cite{Kinnersley:1969zza}, see also e.g. \cite{2012PhRvD..86h4020Z} for its properties), whose basis vectors are (expanded in the coordinate basis $\{dt,dr,d\theta,d\phi\}$)
\bea \label{eq:Tetrad}
l_a &=& \left(1, -\frac{\Sigma}{\Delta},0, -a \sin^2\theta\right)
\,, \notag \\
n_a &=& \frac{1}{2\Sigma}\left( \Delta,\Sigma, 0, -a \Delta \sin^2 \theta \right) 
\,, \notag \\
m_a &=& \frac{1}{\sqrt{2}\Sigma}\big( a\sin\theta(a \cos\theta + ir), 0, -\Sigma(r-ia\cos\theta), 
\notag \\
&&
-(r^2+a^2)\sin\theta (a\cos\theta+ir) \big)\,.
\eea
Under this tetrad, the Faraday tensor can be decomposed into three complex Newman-Penrose scalars $\Phi_0$, $\Phi_1$ and $\Phi_2$, representing the ingoing wave, Coulomb background and outgoing wave pieces of the electromagnetic field, respectively, with the reconstruction formula being
\bea \label{eq:Reconst}
F_{ab} &=& 4 \Re \Phi_1 n_{[a}l_{b]}+4\Im \Phi_1 m_{[a}\bar{m}_{b]} \notag \\
&&+4 \Re(\Phi_2 l_{[a}m_{b]})+4\Re(\Phi_0 \bar{m}_{[a} n_{b]})\,.
\eea
The magnetic flux across the horizon at $r=r_+$ is simply the integral (\cite{Bicak85})
\bea
\int_{\rm EH} F_{\theta \phi}|_{r=r_+} d\theta d\phi \,,
\eea
and the local flux density is thus essentially $F_{\theta \phi}|_{r=r_+}$ (aside from a $r_+^2 \sin\theta$ factor from the integration measure). 
An examination of the Meissner effect is then a computation of $F_{\theta \phi}|_{r=r_+}$ (when we and the broader literature talk about vanishing flux in this context, it is really the flux density, not the total integral through the entire horizon, that is intended). 

Although we have used the Boyer-Lindquist coordinates for concreteness, we note that the tetrad is a geometrical construct that can be used in other coordinate systems as well. More importantly, $F_{\theta \phi}|_{r=r_+}$ is the same in Boyer-Lindquist as in Kerr coordinates (\cite{Bicak76}), so we are at liberty to compute it in the Boyer-Lindquist system, which is more convenient, but the result will nevertheless apply to horizon penetrating coordinate systems. 

In the extremal limit, $r_-$ and $r_+$ become identical and $x=\Delta^{1/2}=r-r_+$ becomes a measure of the distance to the horizon. To study whether the horizon flux density vanishes, one can then examine the leading order (lowest power) dependence of $\Phi_0$, $\Phi_1$, $\Phi_2$ and the tetrad basis on $x$, and in turn that of $F_{\theta\phi}$ through Eq.~\eqref{eq:Reconst}. The relevant tetrad scaling behaviour can be read off directly from Eq.~\eqref{eq:Tetrad}, which are
\bea
&&l_{\theta} = 0\,,  \quad
l_{\phi} \sim x^0\,, \quad
n_{\theta} = 0\,,  \quad
n_{\phi} \sim x^2\,, \notag \\
&&m_{\theta} \sim x^0\,, \quad
m_{\phi} \sim x^0\,,
\eea 
so in order to achieve a non-vanishing horizon flux density, we need any of the three complex scalars to drop at or slower than 
\bea \label{eq:Conds}
\Phi_{0}\sim x^{-2}\,, \quad
\Phi_{1}\sim x^{0}\,, \quad
\Phi_{2}\sim x^{0}\,. 
\eea
Note that because of the singular behaviour of the Kinnersley tetrad on the horizon, $\Phi_0$ is allowed to diverge without causing $F_{ab}$ to also become singular. 
On the other hand, the condition for $\Phi_2$ conflicts with the horizon regularity conditions (see e.g. \cite{Brennan:2013jla})
\bea \label{eq:RegConds}
\Phi_{0}: \mathcal{O}(x^{-2})\,, \quad
\Phi_{1}: \mathcal{O}(x^{0})\,, \quad
\Phi_{2}: \mathcal{O}(x^{2})\,,
\eea
so it won't ever be satisfied, i.e. no wave can come out of the black hole. 

The generality of the Meissner effect as revealed by \cite{Bicak85} boils down to the conclusion that the stationary axisymmetric solutions obey 
\bea \label{eq:Scaling}
\Phi_0 \sim x^{l-1}\,, \quad \Phi_1 \sim x^{l}\,, \quad \Phi_2 \sim x^{l+1}\,. 
\eea
So for $l\geq 1$, $F_{\theta\phi}$ vanishes on the horizon. 
To get this result, one solves the vacuum Teukolsky equation in the extremal and stationary limit, getting 
\bea \label{eq:StationaryPhi2}
\Phi_2 = \rho^2 \sum_{l,m} \,\,{}^2y_{lm} e^{im\phi}\,\,{}_{-1}S_{lm}(\theta)\,,
\eea
with $\rho^2 = M^2/(r-i a \cos\theta)^2$ and ${}_{-1}S$ being the spin-weighted spheroidal harmonics (which are simply the spin-weighted spherical harmonics in the stationary case). The important term in Eq.~\eqref{eq:StationaryPhi2} is 
\bea \label{eq:Phi2}
{}^2y_{lm} = e^{-im/x} x^{l+1} \xi\left(1-l,-2l;\frac{2im}{x}\right)\,,
\eea
where $m$ is the azimuthal quantum number and $\xi$ is the confluent hypergeometric function \footnote{We note that $l=1$ corresponds to a pole of the confluent hypergeometric function $\xi$. Nevertheless, taking the limit of $l \rightarrow 1$ from other nearby values, we see that the solution remains valid in this case, so this result of \cite{Bicak85} is valid for $l\geq 1$.}. After enforcing axisymmetry by setting $m=0$, both the exponential phase factor and $\xi$ in Eq.~\eqref{eq:Phi2} reduce to constants, and we recover the scaling in Eq.~\eqref{eq:Scaling} for $\Phi_2$. 
And to get to the other Newman-Penrose scalars from $\Phi_2$, derivatives against $x$ are taken, at most once for $\Phi_1$, twice for $\Phi_0$, thus the sequence of powers in Eq.~\eqref{eq:Scaling}. 

One immediately see a way to evade the Meissner effect, namely through the breaking of axisymmetry. When $m\neq 0$ is allowed, the confluent hypergeometric function becomes a polynomial up to $1/x^{l-1}$, so as far as the lowest power in $x$ is concerned, all the different $l$ choices contribute at $x^2$ in Eq.~\eqref{eq:Phi2}, rather than starting at higher powers of $x$ (thus more ignorable) for higher $l$'s, as in the $m=0$ case. 
This however, is insufficient in itself for creating a non-vanishing flux at extremality, as $x^2$ is at the same order as the $l=1$ contribution when $m=0$, and we have seen that a dipole is not capable of creating non-vanishing fluxes in that situation. The key is the activation of the exponential phase term $e^{-im/x}$. When we hit $e^{-im/x} x^p$ with $\partial_x$, we get by product rule a term $\propto (x^{p}/x^2) e^{-im/x}$ when the derivative acts on the exponential, this gives a $x^{p-2}$ rather than a simple $x^{p-1}$ as a derivative on the polynomial part would otherwise produce, which makes it possible to create a scaling with sufficiently low power of $x$ to satisfy Eq.~\eqref{eq:Conds}. For example, with $l=3$, $m=3$, we have
\bea
{}^2 y_{33}&=& e^{-3 i/x} \left(-\frac{36}{x^2}+\frac{60 i}{x}+30\right) x^4\,, \\
\frac{d\,\, {}^2 y_{33}}{dx} &=& 6 e^{-3 i/x} \left(20 x^3+45 i x^2-42 x-18 i\right)\,,  \notag \\
\frac{d^2\,\, {}^2 y_{33}}{dx^2} &=& \frac{18}{x^2} e^{-3 i/x} \left(20 x^4+50 i x^3-59 x^2-42 i x+18\right)\,, \notag
\eea
so the first derivative, and subsequently $\Phi_1$, contains an $x^0$ order term, while the second derivative, and subsequently $\Phi_0$, contains an $x^{-2}$ order term, both capable of giving rise to a non-vanishing horizon flux density (see Eq.~\eqref{eq:Conds}). This is technically why the Meissner effect only turns up in axisymmetry. 

Finally, it is interesting to note that $l=0$ is not included in the derivation above and can serve as another way to evade the Meissner effect (\cite{Penna:2014aza}). This corresponds to a magnetic monopole, which can be made more physical by placing a plasma-supported current sheet on the equatorial plane (so the magnetic field does not need to be continuous across it), splitting the spacetime into two halves with magnetic field configurations corresponding to oppositely charged monopoles on either side (the so-called split monopole configuration of \cite{1977MNRAS.179..433B}). It is however worth noting that in the case of broken axisymmetry (and broken stationarity below), $F_{\theta\phi}$ from both $\Phi_0$ and $\Phi_1$ contributions can be non-vanishing on the horizon, while in the monopole case, only that from $\Phi_1$ is allowed to do so (the lowest multipole order for electromagnetic waves is dipole, so there is no ingoing monopolar wave). 
This means that if $\Phi_0$ as the ingoing wave is required for the energy extraction process (e.g. as the carrier of negative energy down the black hole), then the split-monopole configuration may not suffice (note that although one of the original solutions in \cite{1977MNRAS.179..433B} is called a split-monopole solution, because it is obtained by perturbing a monopole in Schwarzschild, the actual energy-extracting solution for slowly spinning Kerr that came out in the end has a non-vanishing $\Phi_0$).

\section{Evasion through dynamic evolution}\label{sec:TimeDep}

Dependence on the azimuthal coordinate $\phi$ is not the only way to produce an exponential phase term. In the Kerr spacetime, the temporal dimension has a fair amount of similarities with the azimuthal direction. For example, $\partial_t$ and $\partial_{\phi}$ are both Killing vectors of the underlying spacetime (the electromagnetic field does not need to possess all of the symmetries of the spacetime though, so it can break either axisymmetry or stationarity, or both), and both $t$ and $\phi$ enter through simple exponential terms in the separation of variables expression for $\Phi_2$ \footnote{As we are examining the effects of the fast oscillatory temporal variations here, and not the secular growth or decay that requires an accompanying energy change in the electromagnetic field, we concentrate on energetically steady evolutions with $\omega \in \mathbb{R}$. Nevertheless, we point out that when $\Im\omega \neq 0$ and if its value is oversimplified as compared to the real astrophysical situation, e.g. when we have a secular decay at constant rate lasting for an unlimited duration, pathological behaviours such as a diverging amplitude in the infinite past may appear, and may inject well-known unphysical pathologies into the horizon limiting properties of $\Phi_2$ depending on the coordinate choice (see \cite{Jacobson:2011ua} and Sec.~II C in \cite{2015PhRvD..91h4007Y}). 
Furthermore, note that as we allow arbitrary currents and charges (including current sheets) outside of the vacuum layer, 
we don't have the restrictive matched asymptotic expansion conditions seen for everywhere vacuum quasinormal modes that quantize $\omega$ (\cite{Teukolsky:1974yv}).}:  
\bea \label{eq:FullSep}
\Phi_2 &=& \rho^2\sum_{l,m} \int d\omega e^{-i\omega t}e^{im \phi}\,\,{}_{-1}S_{lm\omega}(\theta)\,\,{}^2 R_{lm\omega}(r)\,. \eea
Therefore, one is naturally led to the expectation that the breaking of stationarity may also result in an evasion of the Meissner effect. 

The relevant equation now is the time-dependent version of the radial Teukolsky equation (Eq.~4.9 of \cite{Teukolsky})
\bea \label{eq:Teu}
\left[ \frac{K^2+2i(r-M)K}{\Delta}-4i \omega r -\lambda_{lm\omega}\right] \notag \\
+ \Delta \frac{d^2 ({}^2R_{lm\omega})}{dr^2} = 0\,,
\eea
with
\bea
K &=& (r^2+a^2)\omega -a m\,, \notag \\
\lambda_{lm\omega} &=& A_{lm\omega}+a^2 \omega^2 - 2 a m\omega \,,
\eea
where 
\bea \label{eq:Alm}
A_{lm\omega} \approx l(l+1)-\frac{a^2\omega^2}{2}\left(1-\frac{m^2}{l(l+1)}\right)\,,
\eea
when $a^2\omega^2/l(l+1)$ is small (see Eq.~2.14 of \cite{Yang:2013uba}). We will make the small $|\omega|$ assumption below in order to obtain exact solutions, but present numerical experiments and physical reasoning to argue that the important Meissner-effect-evading properties should be preserved for larger $|\omega|$ cases as well. Concentrating on a single constituent under the summation and integral signs of Eq.~\eqref{eq:FullSep}, and dropping the subscripts for brevity, the extremal limit of Eq.~\eqref{eq:Teu} for $m=0$ (we concentrate on evading the Meissner effect without breaking axisymmetry in this section) is explicitly
\bea \label{eq:TimeDepEq}
&&{}^2R(x) \Bigg[x^2 \left(-l (l+1)-\frac{\omega^2}{2}-4 i (x+1) \omega \right)
 \notag \\
&&
+\left((x+1)^2+1\right) \omega  \left(\left((x+1)^2+1\right) \omega +2 i x\right)\Bigg]
 \notag \\
&&
+x^4 \,\,{}^2R''(x)=0\,.
\eea
We immediately notice that the structure of Eq.~\eqref{eq:TimeDepEq} is similar to that of the stationary and $m\neq0$ case (e.g. it has a complex coefficient to ${}^2R$ that is also present in the stationary $m\neq 0$ case, but not the $m=0$ case), which is 
\bea \label{eq:Axi}
x^4 \,\,{}^2R''(x)+{}^2R(x) \left[-l (l+1) x^2+m (m-2 i x)\right]=0\,.\,\,\,\,\,\,\,\,\,
\eea
With Eq.~\eqref{eq:Axi}, a change of variable in the following form 
\bea \label{eq:VarChg}
{}^2 R(x) = e^{-\frac{\zeta }{2}} \eta (\zeta ) \left(\frac{D}{\zeta }\right)^{-l}\,, \quad 
\zeta = D/x,
\eea
with the choice of $D=-2i m$ turns it into a confluent hypergeometric equation
\bea \label{eq:Simp}
-\zeta  \eta ''(\zeta )+(\zeta -2 -2 l) \eta'(\zeta )+l \eta (\zeta )=0\,,\,\,
\eea
whose solution appropriate for the near-horizon region is (\cite{Bicak76})
\bea
\eta = D^{2l+1} e^{\zeta} \zeta^{-(2l+1)}\xi(1-l,-2l;-\zeta)\,,
\eea
where the constant factor $D^{2l+1}$ is introduced for convenience. 
With Eq.~\eqref{eq:TimeDepEq}, the same transformation \eqref{eq:VarChg} with the choice of $D=4i\omega$ leads to 
\bea \label{eq:FullTimeEq}
0&=&\omega  \left[-\zeta  \eta ''(\zeta )+(\zeta -2 -2l) \eta '(\zeta )+l \eta (\zeta )\right] 
\notag \\
&& +2 i \omega ^2 \eta (\zeta )-\frac{(15 \zeta +16) \omega ^3 \eta (\zeta )}{2 \zeta ^2}
-\frac{16 i \omega ^4 \eta (\zeta )}{\zeta ^2}
\notag \\
&&
+\frac{16 \omega ^5 \eta (\zeta )}{\zeta ^3}+\mathcal{O}(\omega^6)
\eea
where we have expanded in the small quantity $\omega$. Keeping to the lowest order in $\omega$, we recover equation \eqref{eq:Simp}. 
Therefore, the solution for the slowly varying time-dependent case is simply
\bea \label{eq:Phi3}
\Phi_2 &=& \rho^2 \sum_{l} \int d{\omega} e^{i\omega(2/x-t)} x^{l+1} 
\,\,{}_{-1}S_{l0\omega}(\theta)\times \notag \\
&& \times
\xi\left(1-l,-2l;\frac{-4i\omega}{x}\right) \,,
\eea
whose $x$ dependent terms are the same as those in Eq.~\eqref{eq:Phi2} with the simple replacement of $m \rightarrow -2 \omega$, and so 
exactly the same mechanism for evading the Meissner effect applies. Specifically, the confluent hypergeometric function boosts the scaling contribution to $\Phi_2$ from all $l\geq 1$ multipoles from $x^{l+1}$ to $x^2$, and then the phase factor contributes additional $1/x^2$ boosts once derivatives against $x$ are taken, allowing $\Phi_1$ and $\Phi_0$ to satisfy condition \eqref{eq:Conds}. 

\begin{figure}[t,b]
\centering
\begin{overpic}[width=0.99\columnwidth]  {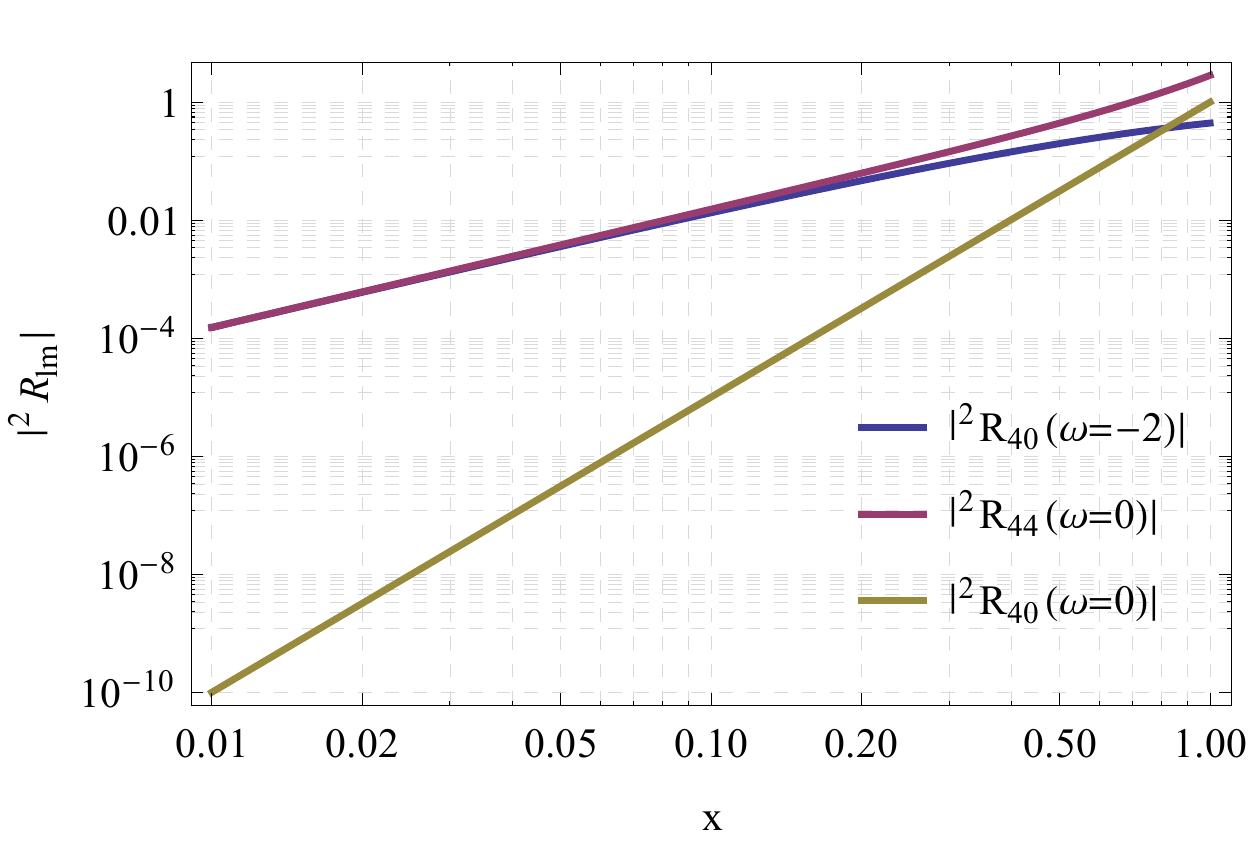}
\put(90,0){(a)}
\end{overpic}
\begin{overpic}[width=0.99\columnwidth,right]  {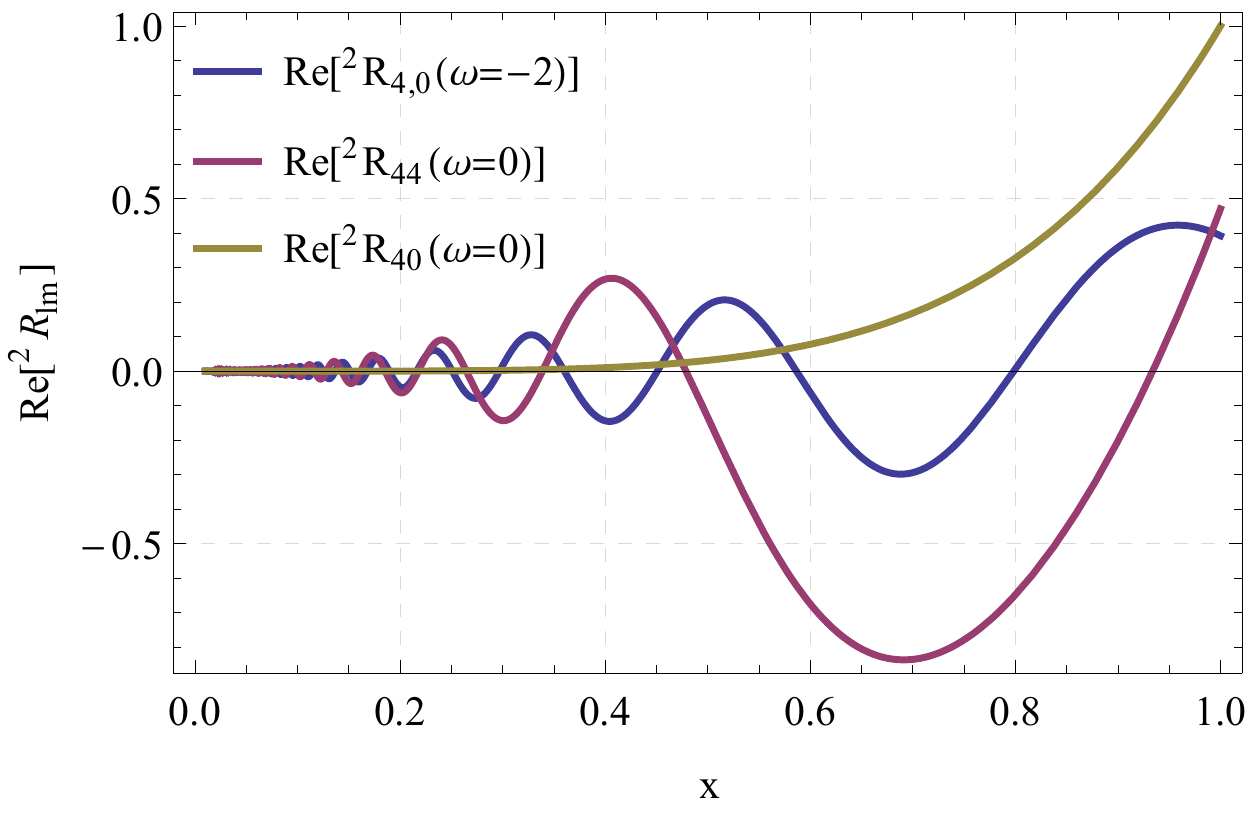}
\put(90,2){(b)}
\end{overpic}
\begin{overpic}[width=0.95\columnwidth,right]  {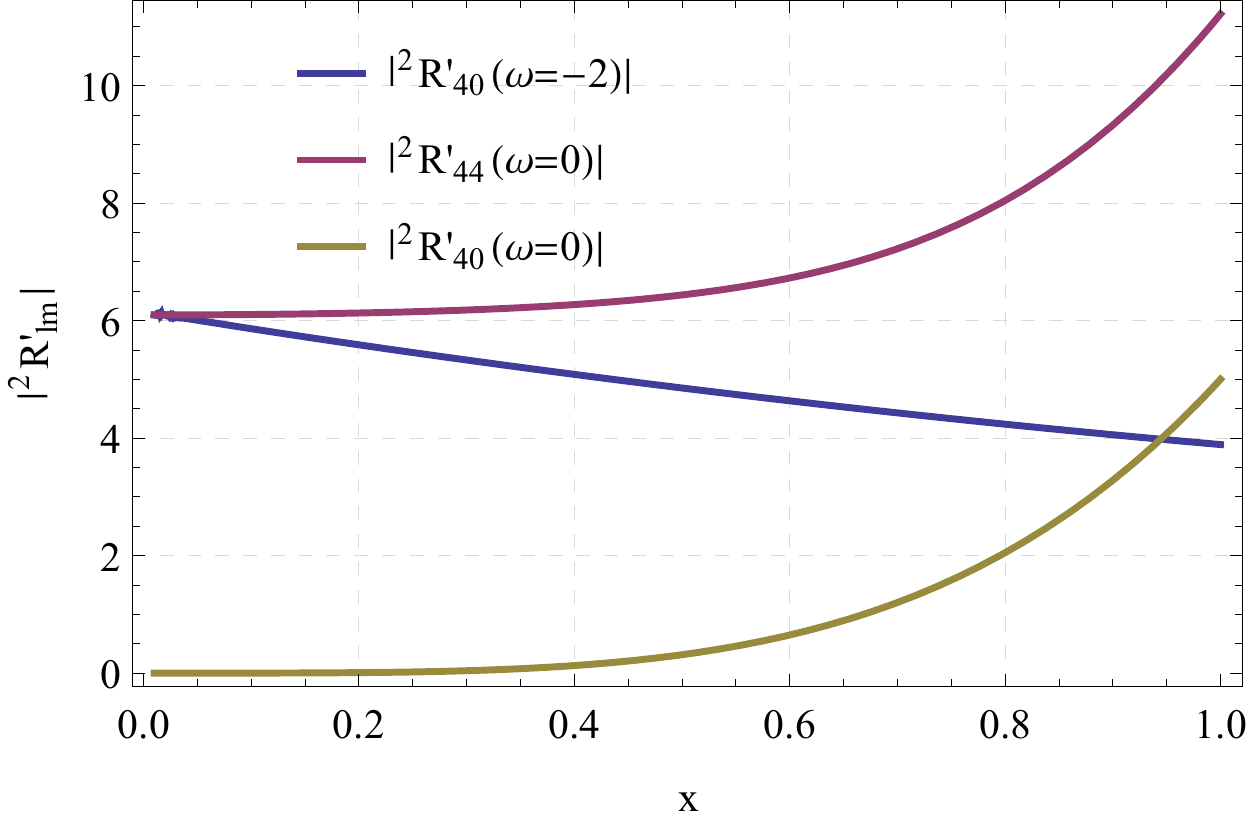}
\put(90,2){(c)}
\end{overpic}
  \caption{
The radial component ${}^2R$ of $\Phi_2$ for the cases with and without axisymmetry and stationarity (the solution to the time-dependent equation \eqref{eq:TimeDepEq} is obtained numerically).   
(a) In this log-log plot, the slope of the lines corresponds to the power indices of power law scalings. 
For $1 < l=4$, the two stationary solutions do not share the same scaling behaviour for $|{}^2R|$. The axisymmetric case with $m=0$ scales as $x^{5}$, while the rotationally asymmetric case with $m=4$ has a scaling of $x^2$. The blue curve at small $x$ values coincides with, and is covered by, the purple one. Therefore the time-dependent $|{}^2R|$ also scales as $x^2$.   
(b) Plotted are the $\Re({}^2R)$ values. The breaking of either axisymmetry or stationarity produces oscillations with increasing frequency as we approach the horizon, symptomatic of a $1/x$ dependent phase factor. 
(c) First derivative of ${}^2R$ shows that, just like the $m\neq 0$ case, the time-dependent solution has a non-vanishing ${}^2R'$ in the horizon limit. 
}
	\label{fig:TimeDep}
\end{figure}

To probe the large $|\omega|$ behaviour of Eq.~\eqref{eq:TimeDepEq} (and to provide a visual example), we now turn to numerical solutions. We pick $l=4$ and $\omega=-2$ for demonstration, so that $a^2\omega^2/l(l+1)=1/5$ is still small and Eq.~\eqref{eq:Alm} can be used, but higher powers of $\omega$ itself are no longer negligible. 
The numerical solution to the full Eq.~\eqref{eq:TimeDepEq} (without ignoring higher powers of $\omega$) with these parameters, together with analytical (from Eq.~\eqref{eq:Phi2}) stationary $m=0$ and $m\neq 0$ solutions as references, are plotted in Fig.~\ref{fig:TimeDep}.
To obtain the numerical solution, we need to specify ${}^2R(x)$ and ${}^2R'(x)$ values at some $x=x_0$ as boundary conditions, and then integrate Eq.~\eqref{eq:TimeDepEq} away from there.
The criteria for the admissibility of any particular boundary condition choice is that the resulting ${}^2R$ should satisfy the horizon regularity condition \eqref{eq:RegConds}, or in other words, $|{}^2R|$ should scale as $\mathcal{O}(x^2)$. Looking at Eq.~\eqref{eq:FullTimeEq}, we notice that the higher powers of $\omega$ tends to be accompanied by higher powers of $x$. So at very small $x$ values, one expects that even in the case of large $|\omega|$, the solutions to Eq.~\eqref{eq:FullTimeEq} may be reasonably approximated by Eq.~\eqref{eq:Phi3}. Therefore, we can try using Eq.~\eqref{eq:Phi3} to set the boundary conditions at some small $x_0$. We note that $x_0$ cannot be pushed all the way to the origin, as Eq.~\eqref{eq:FullTimeEq} is singular at $x=0$ due to the coefficient of the highest derivative term in it vanishing there. This also bleeds numerical issues into the exceedingly small $x$ regime ($x<0.01$), making the acquisition of a reliable numerical solution difficult there (e.g. violent and unpredictable dependence on the working precision of the numerical routine is seen). Nevertheless, there is no problem with picking $x_0=0.01$ and then integrating away from the horizon. The solution thus obtained turns out to be reasonable, demonstrating 
a horizon-regularity-consistent $x^2$ scaling for $|{}^2R|$ (Fig.~\ref{fig:TimeDep}(a)). It also contains a phase oscillation with increasing frequency as we approach the horizon (Fig.~\ref{fig:TimeDep}(b)). Furthermore, after a numerical derivative, ${}^2R'$ is explicitly seen to scale as $x^0$ (Fig.~\ref{fig:TimeDep} (c)), just as in the stationary but rotationally asymmetric case. 

Our numerical experiment thus suggests that the important Meissner-effect-evading ingredients of ${}^2R$ are not spoiled by large $|\omega|$. Aside from the technical observation that Eq.~\eqref{eq:Phi3} appears to provide reasonable approximate large $|\omega|$ solutions in the important small $x$ regime 
(so a better -- but not rigorously proven -- characterization for Eq.~\eqref{eq:Phi3} is perhaps that it is valid when either $|\omega|$ or $x$ is small), we can also make physical arguments as to why it should be so.  
For example, the $x^2$ scaling for $|{}^2 R|$ even when $l>1$ is not surprising as it is the minimal requirement for the horizon regularity condition \eqref{eq:RegConds} to be satisfied, so it is the least special/restricting and should be generic among field configurations that are not severely constrained by e.g., multiple symmetries. Equivalently stated, further restricting the field to satisfy \eqref{eq:RegConds} beyond the minimal order is unnecessary and wasteful as far as horizon regularity is concerned, so such restrictions, if exist, have to come from other considerations.   
More importantly, the physical meaning of the oscillatory phase factor in the time-dependent case is even more apparent than in the rotationally asymmetric case. This factor simply enforces an ingoing-wave boundary condition at the event horizon as demanded by causality
\bea 
{}^2R \propto e^{-i\omega(t+ r_*/v)}\,,
\eea
where $v>0$ is the radial velocity, 
while $r_*$ is the tortoise coordinate that can be seen as a function of $x$ defined by $dr_*/dx=(r^2+a^2)/\Delta$, which is approximately $2/x^2$ near the horizon, and so $r_* \sim -2/x$ with an appropriate integration constant. 
Specifically for a rotating black hole, ${}^2 R$ asymptotes to (see e.g., \cite{Yang:2013uba})
\bea \label{eq:Res}
{}^2R\propto e^{-i(\omega -m\Omega_H)r_*} e^{-i\omega t}\,,
\eea
on the horizon, where $\Omega_H$ is the horizon angular velocity. Setting $m=0$, Eq.~\eqref{eq:Res} gives us the phase factor in Eq.~\eqref{eq:Phi3}. In addition, because the ingoing-wave condition \eqref{eq:Res} is generic for all solutions of the original unsimplified radial Teukolsky equation, the same factor should also appear in ${}^2R$ even when $|\omega|$ is large. On the other hand, setting $\omega=0$ and noting that $\Omega_H=1/2$ at extremality, Eq.~\eqref{eq:Res} reduces to $e^{-im/x}$, agreeing with Eq.~\eqref{eq:Phi2} for the stationary and rotationally asymmetric case discussed in Sec.~\ref{sec:Review}.  

\section{Mode coupling via energy injection}\label{sec:Turbulence}

\begin{figure}[t,b]
\begin{overpic}[width=0.47\columnwidth]  {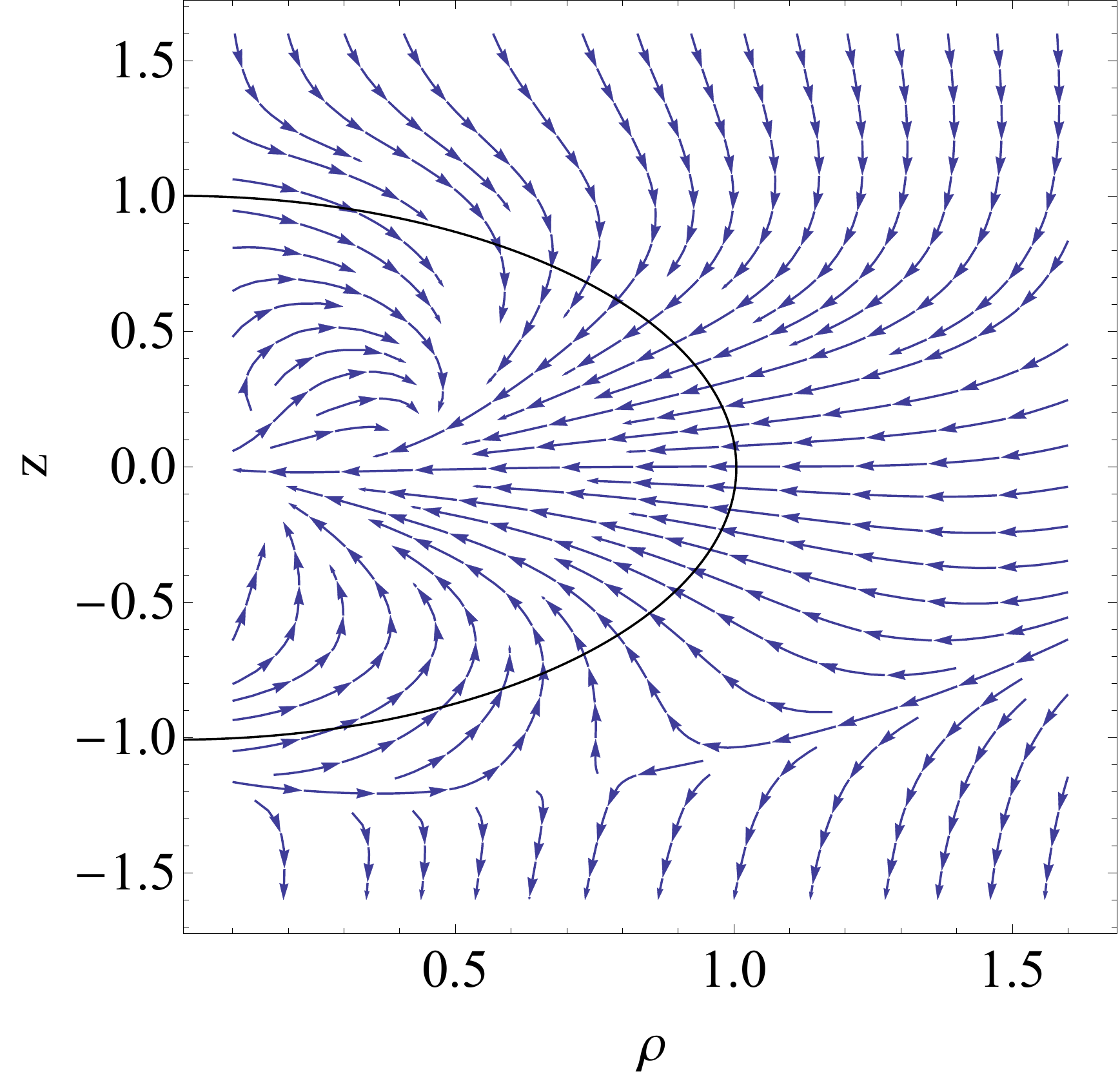}
\put(90,0){(a)}
\end{overpic}
\begin{overpic}[width=0.47\columnwidth]  {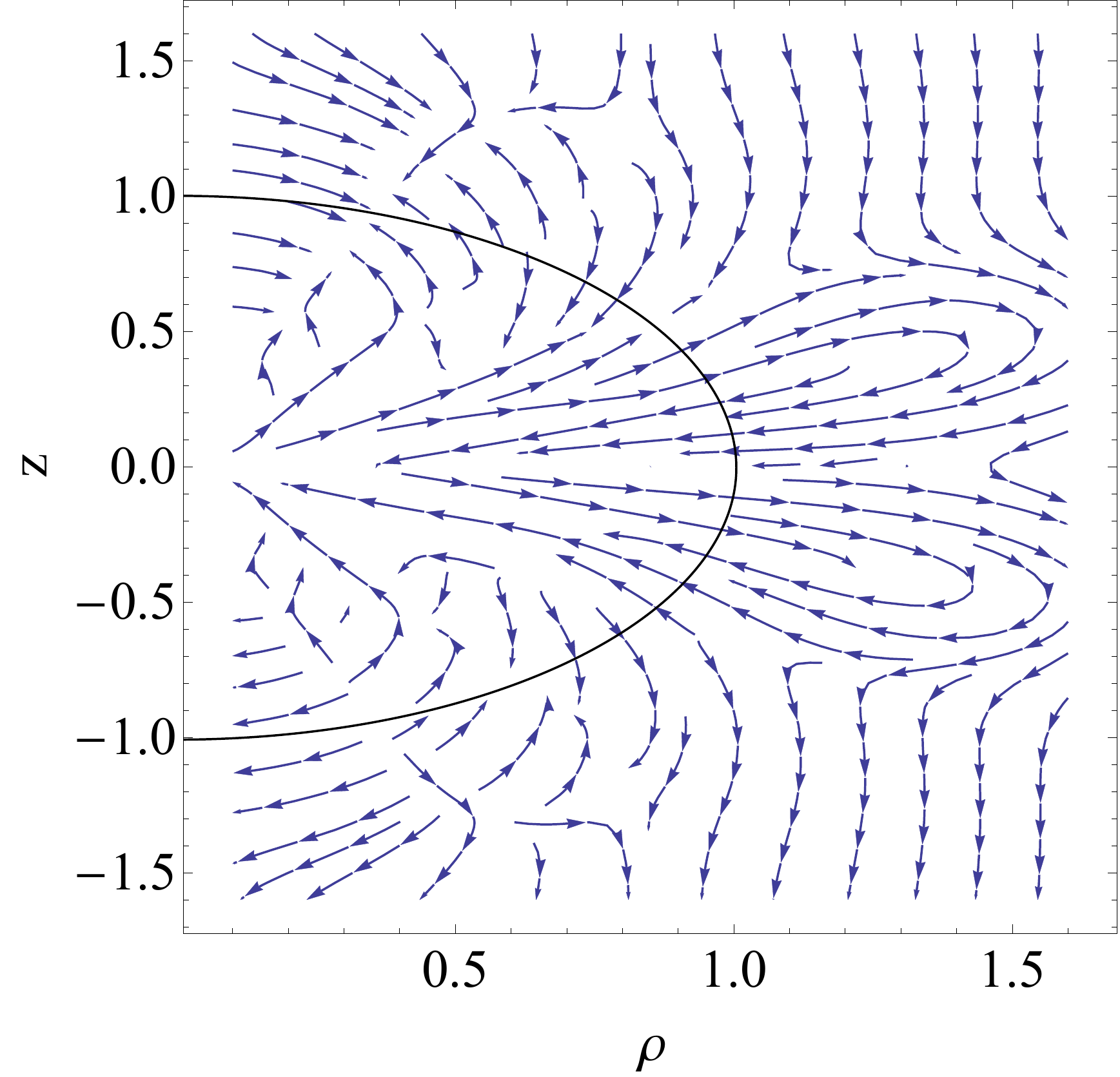}
\put(90,0){(b)}
\end{overpic}
  \caption{The perturbed magnetic field lines in the cylindrical counterpart to the Ingoing Kerr coordinates, and with $\tilde{t}=0$, $\omega=0.01$, and $a=0.99999$: 
  (a) Adding an $(l=1, m=0)$ mode $\Phi_0$ to the Wald solution introduces a change in the magnetic field line directions, which now thread through the horizon while originally flowing around it as in Fig.~\ref{fig:WaldStream}(c). 
  (b) Higher $l=10$ perturbing mode in $\Phi_0$ also causes the field lines to thread through the horizon. However, more lines become closed, emerging and returning to the horizon, rather than extending out to infinity. 
}
	\label{fig:FluxUnderPert}
\end{figure}

Within our discussion, force-free currents, which are nonlinear functions of the electric and magnetic fields, should be present outside of the boundary layer, and have the potential to bring about coupling between time-dependent perturbations characterized by different angular quantum numbers $l$ (abbreviated to ``modes" below), and induce turbulence when given the right conditions. 
Analytical studies (\cite{1995ApJ...438..763G}) and numerical simulations (\cite{2005ApJ...621..324C,ZrakeEast2015}) in the flat spacetime show that force-free turbulence possesses a Kolmogorov energy spectrum with power index $-5/3$. However, for freely decaying MHD (not magnetically dominated)  turbulences, \cite{2014ApJ...794L..26Z} and \cite{2015PhRvL.114g5001B} have measured indices closer to $-2$.

Aside from such direct nonlinear couplings, there may be indirect energy injection effects that are also capable of ``coupling" the modes, in the sense that a single seeding mode can excite siblings. Essentially, the spherical harmonic content of the electromagnetic perturbation appearing in the Newman-Penrose scalars are not faithfully reflected in the horizon flux density, because in order to reconstruct the Faraday tensor according to Eq.~\eqref{eq:Reconst}, one has to involve the Newman-Penrose basis vectors which are themselves functions of $\theta$. For concreteness and some variety in coordinate choice, let's consider Ingoing-Kerr coordinates 
$(\tilde{t},r,\theta,\tilde{\phi})$, which share the same $r$ and $\theta$ with the Boyer-Lindquist coordinate system, while the relationship for the temporal and azimuthal coordinates are 
\bea \label{eq:CoordTrans}
\tilde{t}= t + r_* - r\,, \quad 
\tilde{\phi} = \phi + \frac{a}{r_+-r_-} \ln \left|\frac{r-r_+}{r-r_-}\right|\,.
\eea 
The metric is now given by 
\bea
ds^2&=& -\left(1-\frac{2 M r}{\Sigma }\right)d\tilde{t}^2
+H^2 dr^2 
+\Sigma d\theta^2 \notag \\
&&+\frac{A \sin ^2\theta }{\Sigma }d\tilde{\phi}^2 
-2 a  H \sin ^2\theta dr d\tilde{\phi}
\notag \\
&&-\frac{ 4 a M r \sin ^2\theta }{\Sigma }d\tilde{t} d\tilde{\phi}
+\frac{4 M r}{\Sigma }dr d\tilde{t} \,,
\eea
where
\bea
A &=& (r^2+a^2)^2-a^2\Delta \sin^2\theta\,, \notag \\
H &=& \sqrt{1+\frac{2Mr}{\Sigma}}\,. \quad 
\eea
Using Eq.~\eqref{eq:CoordTrans}, it is straightforward to compute the Jacobian for the coordinate transformation and subsequently the expressions for the Kinnersley tetrad in the Ingoing Kerr coordinates; and then using Eq.~\eqref{eq:Reconst}, we obtain that for an, e.g., $\Phi_0$ perturbation to the Wald solution, the horizon flux density is given by (setting $M=1$ for brevity) 
\bea
F_{\theta \tilde{\phi}}=2 \sqrt{2} \Re\left(\frac{((r-2) r+16) \sin ^2\theta }{r-4 i \cos \theta }\Phi_0\right)\,,
\eea
where the extra $\theta$ dependent term multiplying onto $\Phi_0$ makes it clear that even when $\Phi_0$ contains a single solitary $l$ contribution, the horizon flux density will in general contain both higher and lower $l$'s. 

Fundamentally, both the Newman-Penrose scalars and $F_{\theta \tilde{\phi}}$ are components of the Faraday tensor, but under different tetrads, with the former being associated with the Kinnersley tetrad and the latter under the coordinate tetrad $\{d\tilde{t},dr,d\theta,d\tilde{\phi}\}$. The $\theta$ dependent transformation between the tetrads then causes a mismatch between the harmonic decompositions of the tensor components in these respective bases. This mismatch has physical consequences. As the Maxwell equations are separable only in the Kinnersley tetrad, 
it is the $l$ of the Newman-Penrose scalars that should be used to label electromagnetic evolution modes. On the other hand, $P_{\rm BZ}$ is proportional to flux squared, so it supplies energy according to the harmonic components in $F_{\theta \tilde{\phi}}$. There is no reason to expect this alteration in harmonic contents to get exactly reversed when the extracted energy is fed back into the electromagnetic variation modes, i.e. for the energy to simply proportionally enhance the original seeding mode. Although a rigorous analytical proof for this energy extraction detail is not currently available \footnote{\label{fn:Sqr} Even the stationary monopole-like BZ solution for rapidly-spinning black holes is still not known analytically, although its power should remain proportional to magnetic flux density squared on dimensionality grounds, as even though correction terms with higher powers of $a$ may appear, $a$ is itself dimensionless, so we still need magnetic field squared to create an energy density-like coefficient.}, we note that for the extracted energy to leave the horizon, outgoing $\Phi_2$ must be involved even when the seed field is purely ingoing containing only $\Phi_0$, so the field configuration has to change more than just an overall amplitude. 
Therefore, energy most likely gets pumped into other sibling modes that may not be present initially. 
This sequence then repeats, and energy cascades and inverse-cascades into a broad range of available length scales, eventually establishing a steady turbulent spectrum. 

\begin{figure}[t,b]
\centering
\begin{overpic}[width=0.9\columnwidth]  {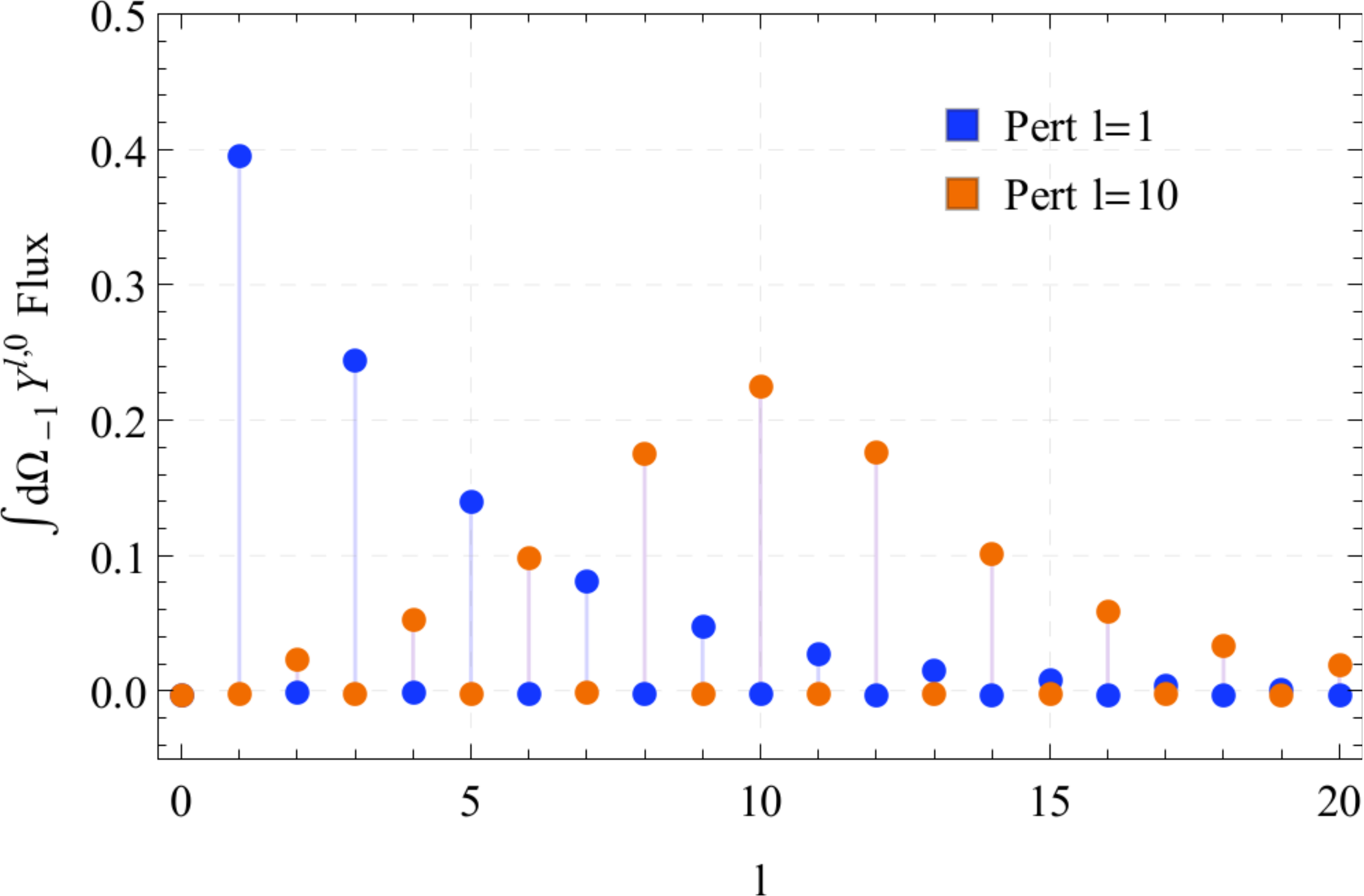}
\end{overpic}
  \caption{
The innerproduct of the flux density across horizon (resulting from an $l=1$ or $l=10$ seed perturbation in $\Phi_0$ as depicted in Fig.~\ref{fig:FluxUnderPert}) with spin-weighted spherical harmonics ${}_{-1}Y^{l,0}$ (appropriate for $\Phi_2$ that carries energy outwards). A significant presence of higher $l$ modes are seen in the flux even when the initial perturbation contains only $l=1$. With initial perturbation of a higher $l=10$, harmonic contributions leak into both higher and lower $l$'s, in a more symmetric fashion. The innerproduct values are normalized so contributions from all the modes for say, the $l=1$ seed, add up to unity. 
}
	\label{fig:lComp}
\end{figure}

For a more visual demonstration, consider an ingoing perturbation in $\Phi_0$ of the simple toy form 
\bea \label{eq:Toy}
\Phi_0 &=& \frac{1}{2}\frac{e^{-i\omega(\tilde{t}+r)}}{(r-r_+)^2}  \,\,{}_{1}Y^{l,m=0}(\theta, \tilde{\phi}) \,, 
\eea
on top of a background Wald solution, 
where ${}_{1}Y^{l,m}$ are the spin-weighted spherical harmonics (we choose $\omega=0.01$ so $a\omega \ll 1$ and the spherical harmonics are good approximates to the spheroidal harmonics, for whom we do not have simple analytical expressions). 
Note, Eq.~\eqref{eq:Toy} satisfies the ingoing boundary condition \eqref{eq:Res} (noting $\tilde{t}+r=t+r_*$), but does not properly solve the radial Teukolsky equation everywhere (we do not have exact solutions for $\Phi_0$ in the dynamic case). In any case though, it is the angular behaviour that interests us in this section, so the radial part just needs to be sufficiently large near the horizon. We then use Eq.~\eqref{eq:Reconst} to reconstruct the Faraday tensor and plot the perturbed magnetic field lines in Fig.~\ref{fig:FluxUnderPert}. 
We can also estimate the spherical harmonic components in the horizon flux density, which is plotted in Fig.~\ref{fig:lComp} and shows that higher $l$'s are generically present even when the initial perturbation consists of only $l=1$. In this case, there are not many modes with even lower $l$, so the ``leaking" is asymmetric. In contrast, high initial $l$ perturbations generate both lower and higher $l$ components in the flux density fairly symmetrically (although the lower $l$ side is still cut off at $l=0$). As it is reasonable to expect more patchy energy patterns from higher $l$ components in $F_{\theta \tilde{\phi}}$ to be better matched and thus absorbed by corresponding higher $l$ modes of the electromagnetic field, this flux density harmonic decomposition pattern can be seen as an approximate surrogate for the mode excitation pattern, and the overall trend is a spreading of energy among different $l$'s. 

Although this energy injection effect looks fairly different from the usual direct mode-couplings (if we view our modes as driven oscillators, then they are coupled through the driving force's dependence on other modes), it nevertheless represents a robust and efficient mechanism for cascading and inverse-cascading energy throughout the entire spectrum, exciting previously absent modes to establish rich angular structures in the electromagnetic field that evolve with even richer dynamics. 
On the other hand, we also have dissipative processes that are likely more prominent in a high $l$ environment \footnote{We expect the emergence of a great many current sheets (where electric field accelerates charges, leading to a transfer of the electromagnetic field energy into thermal energy) when the magnetic field lines become highly fragmented, as large curl in ${\bf B}$ encourages growth in ${\bf E}$. This expectation appears to be verified by \cite{ZrakeEast2015}, who saw a reduction in the rate of dissipative losses when energy is transferred to larger length scales.}. 
We do not know of the quantitative details of the energy injection and dissipation processes, or how large the contributions from other coupling mechanisms are in comparison, but if we assume that in a steady turbulent state (in which the aforementioned energy gains and losses are balanced), the energy spectrum is only a function of the variation frequency $\omega$, and the rate of energy dissipation per unit volume, then one can argue for a Kolmogorov $-5/3$ power law dependence on $\omega$ on dimensional grounds. 
In reality, the situation may be more complex, with the power index not being exactly $-5/3$, or that the steady state is only quasi-steady, exhibiting boom-and-bust cycles when the energy injection or dissipation process temporarily overpowers the other. 
Sophisticated numerical simulations are needed to produce a more accurate characterization. 

\section{Observational Implications} \label{sec:Obs}
There are several alternatives for evading the black hole Meissner effect associated with extremal spin: the (split) monopole configuration (if its lack of $\Phi_0$ poses no problem), the breaking of axisymmetry, having the black hole carry electric charge, ITV, and possibly others. We will discuss some potential observational consequences of ITV, but note that other evasion mechanisms would likely be present simultaneously. Of those, the monopole configuration may be the more common, as astrophysical black holes are expected to be electrically neutral, and there is no obvious reason to expect their spin axes to be systematically misaligned with the magnetic field (\cite{Penna:2014aza}). 
We further note that although monopoles are not natural for jet formation, gradual collimation occurring further away from the black hole horizon can possibly compensate (\cite{1999Natur.401..891J}). 

On the other hand, as mentioned in Sec.~\ref{sec:Intro}, we expect ITV to be rather ubiquitous among systems powered by rapidly-spinning holes. In what follows, we will restrict our discussion to AGNs of this character (amounting to most of them, according to the estimates of \cite{Reynolds:2013rva}, \cite{2006ApJ...642L.111W}, and \cite{2002ApJ...565L..75E}). Furthermore, black hole X-ray binaries such as Cygnus X-1 are assumed to share much the same energy generation and emission processes as AGNs, and are as such named microquasars. Our discussion below will apply to microquasars the same way as it does to AGNs, and we will utilize either class of objects depending on available observational data. 

There are a plethora of different AGN categories defined according to their observational signatures, but schemes have been devised to combine them into a single unified picture. One leading contender is based on the orientation of the jets (\cite{1993ARA&A..31..473A,1995PASP..107..803U,1990A&ARv...2..125B}). Specifically, most or all AGNs contain jets, an accretion disk and an axisymmetric dusty torus. The disk material heats up and radiates thermally in the optical to ultraviolet bands. Furthermore, a corona of hot material above the disk can inverse-Compton scatter softer photons upto X-ray energies. The jet also radiates, from radio to gamma-ray bands through synchrotron and inverse-Compton processes, and so the radiations are non-thermal. When a jet is pointed close to our line of sight (e.g. with the case of blazars), we can see much non-thermal radiation produced inside the jet, in addition to those originating from the disk and corona. For example, X-ray continuum emissions from the jet and the corona share a common scattering origin, while radio signals are considered to be from the jet alone. On the other hand, if the jet is pointed elsewhere, thermal radiations from the disk become more prominent, and we also need to be careful about scattering by the dusty torus.

A particularly interesting aspect of AGNs and microquasars is that they exhibit luminosity variations. Despite the current poor understanding of the underlying mechanism for this phenomena (\cite{Wold:2006eb}), it is considered to be of great importance as it may help diagnose the fundamental physics behind the various emission activities (\cite{1994MNRAS.268..305H}). Of the many constituents to the central core of AGNs and microquasars, previous explanations for variability have largely concentrated on the disk, in order to take advantage of its rich dynamics. Black holes on the other hand, are rather unlikely to spontaneously vary, since radiating away its time dependence through quasinormal ringing appears to be the norm. There is however another constituent that's been subjected to less scrutiny: the electromagnetic field threading the black holes, which is quite capable of having its own intrinsic nontrivial dynamic evolution in a curved spacetime 
(the dynamics can initially be triggered by other processes though). In other words, variabilities driven by ITV is an enticing possibility. 

An ITV-based explanation would corroborate with the assertion that variability is an important tool for probing AGN and microquasar energetics, as ITV ties in to the very core of the black hole rotational energy extraction process and directly regulates the extraction efficiency, which also means it is capable of generating the type of large variation amplitudes seen observationally. In addition, features of the electromagnetic field naturally travel at the speed of light, thus capable of creating the extremely short timescales that's on the order of light crossing times. Indeed, it is due to the prevalence of such short timescales (not only in the variations themselves, but also in lags between different wavebands, etc) that \cite{2004ApJ...612L..21G} speculated that the mechanism underlying variations should be electromagnetic in nature (although it is electromagnetic processes in the disk, rather than in the vicinity of the black hole that's envisioned there). In the next three sections, we propose several possible links between ITV and observations, in the context of variability and other related issues.

\subsection{Variability-flux correlation} \label{sec:VarFluxCorr}
Perhaps the most straightforward prediction of ITV is that, because it is invoked to increase the energy extraction efficiency of the BZ process, one can expect AGNs and microquasars to become more luminous during those periods of pronounced variation. Furthermore, as ITV occurs in the core region in the vicinity of the black holes, the relevant time scale is the light crossing time across a Schwarzschild radius (around $1$ hour for a $10^8\text{M}_{\odot}$ super massive black hole), and so we expect that it is the microvariability \footnote{Variability categories are long-term (on a timescale of several years), short term (months to weeks), and micro (hours to minutes) (see e.g., \cite{2008AJ....135.1384G}). One may also distinguish between the longer term and microvariabilities according to whether the bluer-when-brighter feature is seen in the variations (\cite{2008PASJ...60L..37S}), in which case microvariability also includes shorter period multi-day cases. It is this latter more physical definition that we will adopt here.} that's the most relevant for our discussion.

\cite{Uttley:2001qp,Uttley:2005cc,Superina:2008zz,2010JSARA...4....7D,2003MNRAS.345.1271V,2004ApJ...612L..21G} indeed observed a positive correlation between the root-mean-square amplitude of the microvariations and the average flux during the variation periods (not the long term secular brightness, which would be a different study unrelated to the present discussion) for Seyfert galaxies and microquasars in the X-ray band. In particular, \cite{Uttley:2001qp} recorded a strikingly clean positive linear correlation for Cygnus X-1 (spin $>0.95$) (see Fig.~1 therein, reproduced in Fig.~\ref{fig:CorrFitting} below), with a positive intercept on the mean flux axis, corresponding to a variability-independent component amounting to about a quarter of the maximum flux value.  
  
For other scenarios, similar linear correlations are also recorded in blazars 
by \cite{2009A&A...503..797G} and \cite{2010ApJ...718..279G}, again in the X-ray band. However, in this case, the data points are much more scattered (due to the inherent difficulty in detecting such relations in blazars, as they are faint with large Poisson noise and sampled intensely only during flaring events (\cite{2009A&A...503..797G})) and we don't know if the correlation is still linear. 

In the optical bands, \cite{1987AZh....64..465L} observed a linear correlation for Seyfert galaxy NGC 4151, which is also very clean (see Fig.~5 therein) and with a positive intercept on the flux axis. For blazar W2R1926+42, \cite{2013ApJ...766...16E} reports a very strong linear relationship (Fig.~6 therein), and once again a positive flux axis intercept is seen. In the gamma-ray emissions, \cite{2008bves.confE..16D} also observed a linear correlation for the microvariability of blazar PKS 2155-304 (Figs.~7 and 8 therein). 

For the sake of completeness, we note that another way to examine whether there is a positive correlation between fluxes and variation amplitudes is through the shape of the distribution for the latter. Modeling the flux $F$ exhibiting microvariations as a stochastic process
\bea \label{eq:SDE}
\frac{1}{F^p} dF = \mu dt + \sigma dW \,,
\eea
where $\mu$ is a long term drift, $\sigma$ is the short term volatility and $W$ is the Wiener process, then a positive correlation corresponds to $p>0$. In particular, if $p=1$ (linear correlation), the distribution for $F$ will be lognormal (see Fig.~5 in \cite{2013ApJ...766...16E}), while for more general $p>0$, we will have a distribution skewed towards higher $F$, as signified by a positive skewness (third cumulant over the $3/2$ power of the second cumulant). This method would produce more apparent results when the variance $v$ of the data is large. For example, with lognormal distribution, the skewness is 
\bea
\sqrt{e^{v}-1}\left(2+e^{v}\right)\,.
\eea
This means that it may be difficult to obtain reliable results through this type of integral measurements when fluctuation amplitudes are small (a possible contributing factor to the inconclusive or negative results in some studies looking for lognormal distributions, e.g. \cite{2012Ap&SS.342..147M}), in which case an alternative measurement through binning data and plotting the standard deviation versus mean in each bin would likely be more effective, and is indeed adopted by most studies mentioned above. 

We show now that the ITV-induced variability naturally produces such linear-looking correlations. 
Recall that the power from the BZ process is proportional to magnetic horizon flux density squared (see e.g. Eq.~\eqref{eq:BZP} and footnote \ref{fn:Sqr}), and assume that the energy flux reaching Earth is ultimately traced back to a region on the horizon, whose average flux density is approximated by the illustrative toy model (we will generalize later)
\bea
\mathcal{F}(t)=\mathcal{F}_m+\mathcal{F}_{t}f(\cos(\omega t))\,,
\eea
where $\mathcal{F}_m$ is the monopole contribution, and $\mathcal{F}_t$ is the magnitude of the ITV contribution that we can vary to represent different microvariability amplitudes (we can also add in more ITV components with different frequencies, which will not change the qualitative features below). 
The function $f$ provides flexibility to the profile of temporal variation (as it is not necessarily sinusoidal). Specifically, we assign it the simple form 
\bea \label{eq:TimeProfile}
f(x) = \frac{x}{|x|} |x|^{p}\,,
\eea
so when $p=0$, $\mathcal{F}(t)$ jumps between two discrete values, while larger $p$ provides smoother transitions. Now the average flux (over one or more oscillation periods) and root-mean-square (RMS) variation (not RMS of flux, we take away the mean from the flux before the RMS computation, to match \cite{Uttley:2001qp}) are given by 
\bea \label{eq:FluxExp}
F_{\rm avg} &=& \mathcal{C}\left(\mathcal{F}^2_m+
\mathcal{F}^2_t \frac{\Gamma(p+1/2)}{\sqrt{\pi}\Gamma(p+1)}
\right)+F_{\rm c}\,, \notag \\ 
V_{\rm rms} &=& \frac{\mathcal{C}\mathcal{F}_t}{\sqrt{\pi}}\left[
\mathcal{F}^2_t \left(\frac{\sqrt{\pi} \Gamma(2p+1/2)}{\Gamma(2p+1)}-\frac{\Gamma(p+1/2)^2}{\Gamma(p+1)^2}\right)
\right. \notag \\ &&\left.
+\mathcal{F}^2_m \frac{4\sqrt{\pi} \Gamma(p+1/2)}{\Gamma(p+1)}\right]^{1/2}\,,
\eea
where $\mathcal{C}$ is a constant depending on the black hole mass and spin, as well as the latitude ($\theta$) of the region under consideration, $\Gamma$ is the Euler gamma function, and $F_{\rm c}$ represents a steady (time independent) flux contribution from non-BZ processes such as the disk-driven one in \cite{Blandford:1982di}. 

\begin{figure}[t,b]
\begin{overpic}[width=0.99\columnwidth]  {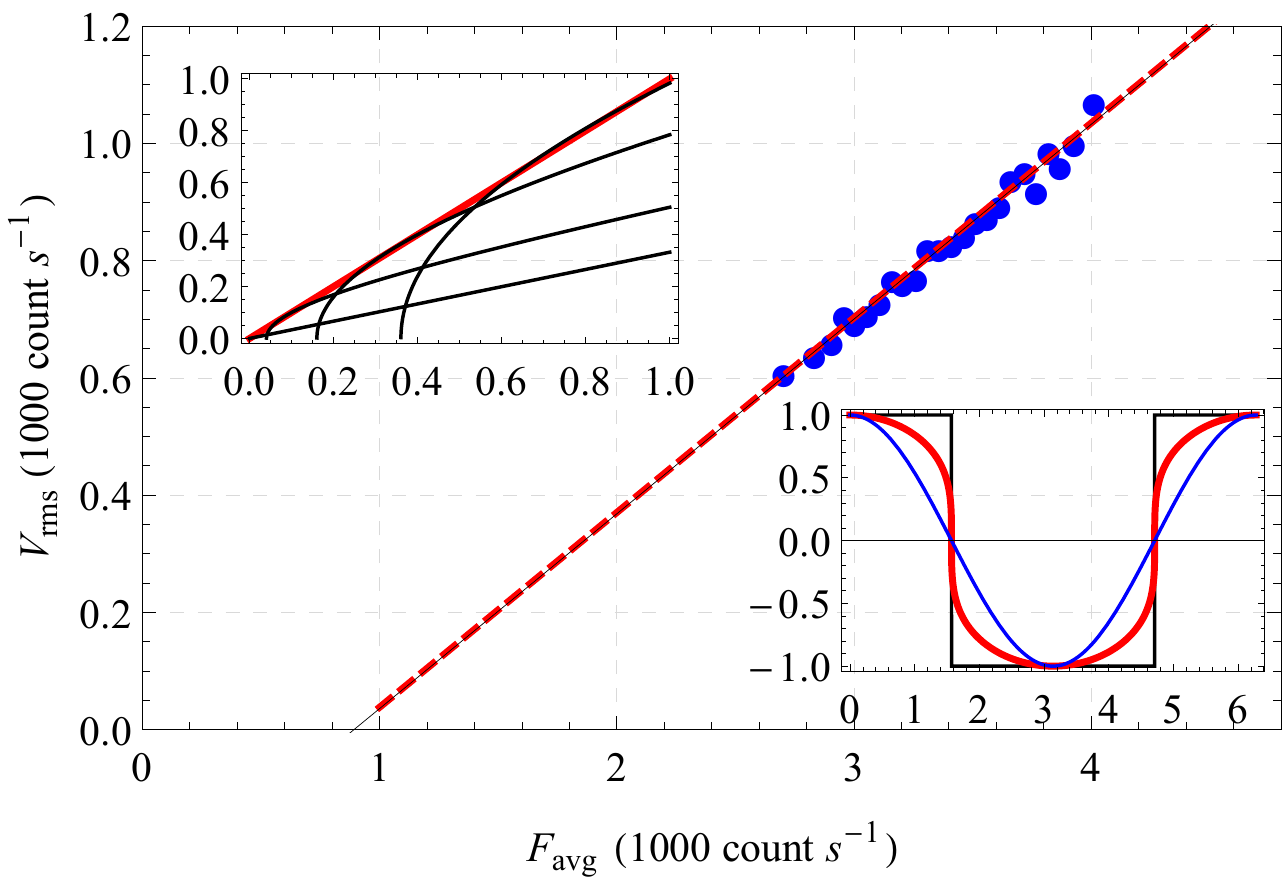}
\end{overpic}
  \caption{
  The mean flux versus root-mean-square variation plot for the Cygnus X-1 microvariability. The blue dots are the observational data, which are taken from Fig.~1 of \cite{Uttley:2001qp}. The dashed red line is the fitting result using Eq.~\eqref{eq:FluxExp}, while the thin black line (beneath the red line) represents the best linear fit. The top-left inset displays curves derived from Eq.~\eqref{eq:FluxExp} with $F_c =0$ and $\mathcal{F}_m \neq 0$. Those associated with larger $\mathcal{F}_m$ values are further to the top-right. The bottom-right inset depicts the magnetic flux density variation profile $f(\cos(t))$ as given by $p$ value choices of $0$ (black), $0.28$ (red, thick, and is the value used for the red line in the main figure), and $1$ (blue). The horizontal axis for the bottom-right inset is time. 
}
	\label{fig:CorrFitting}
\end{figure}

Eq.~\eqref{eq:FluxExp} possesses the following properties that are used to produce the fitting to observational data in Fig.~\ref{fig:CorrFitting}:
\begin{enumerate}
\item The relationship between $F_{\rm avg}$ and $V_{\rm rms}$ is approximately linear (regardless of $p$ choices) when $\mathcal{F}_t$ is large as compared to $\mathcal{F}_m$, but not so when $\mathcal{F}_t$ is small, thus an adjustment to the linear intercept can be expected (see the top-left inset in Fig.~\ref{fig:CorrFitting}, where curves further to the top-right correspond to larger $\mathcal{F}_m$ and show greater adjustments). 

\item The external $F_c$ simply shifts the entire $V_{\rm rms}$ versus $F_{\rm avg}$ curve towards the larger $F_{\rm avg}$ side. On the other hand, the monopole contribution $\mathcal{F}_m$ not only shifts the curve horizontally, but also vertically, so that the curve always touches the $V_{\rm rms} = F_{\rm avg}$ bisection line (see the top-left inset in Fig.~\ref{fig:CorrFitting}). Subsequently, using $\mathcal{F}_m$ alone, we cannot give the linear approximation to the curve a positive intercept on the $F_{\rm avg}$ axis, as is demanded by data. Therefore during fitting, we have used only $F_c$ as a variable and simply set $\mathcal{F}_m=0$, while noting that future observational data in the very low $F_{\rm avg}$ regime is needed to ascertain whether $\mathcal{F}_m$ is in fact present. 

\item The power $p$ in Eq.~\eqref{eq:TimeProfile} determines the slope of the $V_{\rm rms}$ vs $F_{\rm avg}$ curve at large $\mathcal{F}_t$, with $p=0$ corresponding to a vanishing slope, and $p=1$ to a slope of $1/\sqrt{2}$. For the observational data in Fig.~\ref{fig:CorrFitting}, $p=0.28$ provides the best fit. The $f(\cos(t))$ with aforementioned choices of $p$ are shown in the bottom-right inset of Fig.~\ref{fig:CorrFitting}. 

\item There is a degeneracy between $\mathcal{C}$ and $\mathcal{F}_t$, so we simply set $\mathcal{C}=1$ and use $\mathcal{F}_t$ alone for the fitting. The result is the red dashed line in Fig.~\ref{fig:CorrFitting}, which is indistinguishable to the best linear fit (solid black line). 
\end{enumerate}
Note that although we have used an over-simplified deterministic temporal evolution profile for $\mathcal{F}(t)$ for demonstration, the appearance of linear correlation does not depend on this choice. The essence is that the same process for generating the variation is responsible for creating (a part of -- we will ignore $\mathcal{F}_m$ and $F_c$ below for brevity) the mean flux, which is always true with ITV, so both are proportional to the same amplitude parameter, which is $\mathcal{F}^2_t$ in our case. For example, if we let $\mathcal{F}(t)=\mathcal{F}_t g(t)$ with $g(t)^2$ being some fundamental stochastic process in time (more appropriate for a turbulent ITV) so it has fixed mean and variance for each -- and across -- data binning periods of equal duration (different periods can of course differ in their ``volatility" $\mathcal{F}_t$), and let $\langle \cdot \rangle$ represent averaging over a binning period, then we have 
\bea
F_{\rm avg} &=& \mathcal{C} \mathcal{F}^2_t \langle g(t)^2 \rangle\,, \notag \\ 
V_{\rm rms} &=& \mathcal{C} \mathcal{F}^2_t \sqrt{\langle \left[g(t)^2 - \langle g(t)^2 \rangle \right]^2 \rangle} \,.
\eea
So clearly we still have a linear relationship between $F_{\rm avg}$ and $V_{\rm rms}$, with the slope determined by the cumulants of $g(t)^2$ (their counterparts in the deterministic case earlier are adjustable with $p$). 

\subsection{Manifestation in jet emission} \label{sec:Jet}
The jets of AGNs and microquasars are complicated objects involving much as yet unknown physics (\cite{2008JPhCS.131a2052H}), but their polarized non-thermal radiation allows for examination of the magnetic field structure within (\cite{1984RvMP...56..255B,2013EPJWC..6106001W}), to which ITV in the electromagnetic field may be relevant. Indeed, the magnetic field structure of the quiescent (without the shocks of \cite{1979ApJ...232...34B} that are often invoked to explain long term variabilities) jet flow appears to be highly non-trivial. On the one hand, an ordered magnetic field may be required for maintaining the jets' structural integrity (see e.g. \cite{1999ApJ...510..136R}). 
On the other hand, observations such as those on the inhomogeneous linear polarization in radio signals (\cite{1988ApJ...332..678J}) are consistent with there being many randomly oriented magnetic cells (\cite{2005ApJ...621..635H,2011ApJ...735...81H}). 
Therefore, one expects both ordered and random components of the magnetic field to be present. 

The orderliness and randomness are intertwined, but a general trend can be gleaned from observational data. 
The existence of an ordered toroidal component to the magnetic field is supported by rotation-measure gradient records (\cite{2013EPJWC..6106001W}). A possibility is that a largely toroidal field surrounds the jet, stabilizing it against disruptive Kelvin-Helmholtz instabilities through magnetic tension. Such a toroidal field can be produced by currents flowing along the jet and the backflow (Fig.~14 of \cite{1984RvMP...56..255B}).
In the jet spine (middle in the transverse direction), some polarization measurements have demonstrated that the dominant magnetic field direction is along the jet (\cite{2013EPJWC..6106001W}), although unlikely to be unidirectional (\cite{1984RvMP...56..255B}). A possibility is then that inside of the toroidal field and in the jet spine, we have sheared magnetic loops produced by a plasma velocity gradient transverse to the jet direction, stretching loops frozen in the plasma into elongated shapes along the jet, thus enhancing the poloidal fields (see Figs.~12 and 13 of \cite{1984RvMP...56..255B} for a visual depiction). The magnetic field in the spine is otherwise random, as suggested by fractional linear polarization measurements that quantify magnetic randomness. Specifically, the expression for fractional polarization $\alpha$ is (\cite{1994ApJ...437..122W,2013EPJWC..6106001W})
\bea
\alpha=\frac{3 \iota^2 \sin^2\theta'}{2+3 \iota^2 \sin^2\theta'}\,,
\eea
where $\iota$ is the ratio of the uniform over random magnetic field strengths, and $\theta'$ is the angle between jet and observer in the rest frame of the jet material. Using such measurements, \cite{2008ApJ...681L..69G} showed that the magnetic field is indeed quite disordered ($\alpha \sim 5\%$) in the central spine region of the jets (and more ordered at the edges with $\alpha \sim 15-25\%$). 

\begin{figure}[t,b]
\begin{overpic}[width=0.53\columnwidth]  {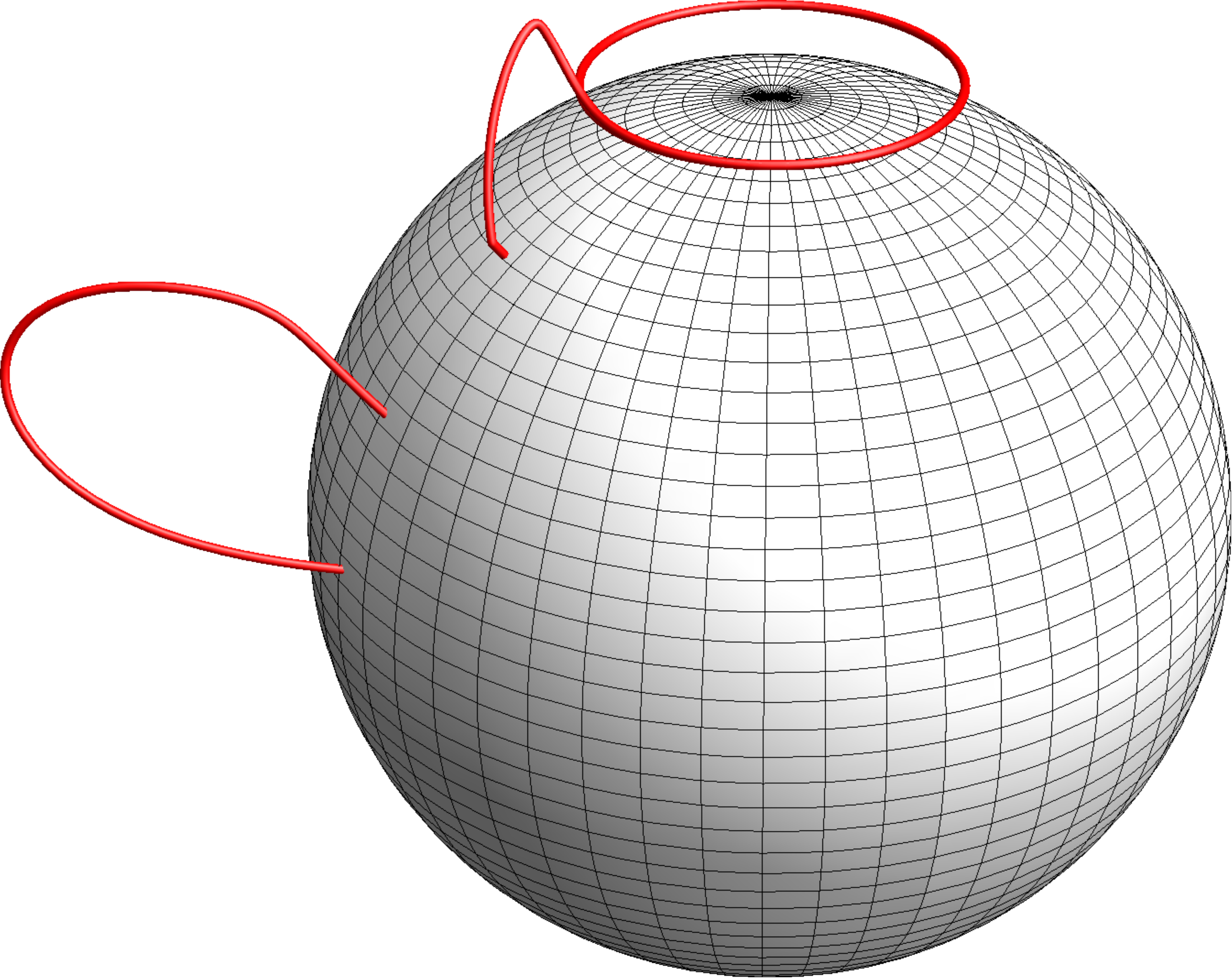}
\put(90,1){(a)}
\end{overpic}
\begin{overpic}[width=0.41\columnwidth]  {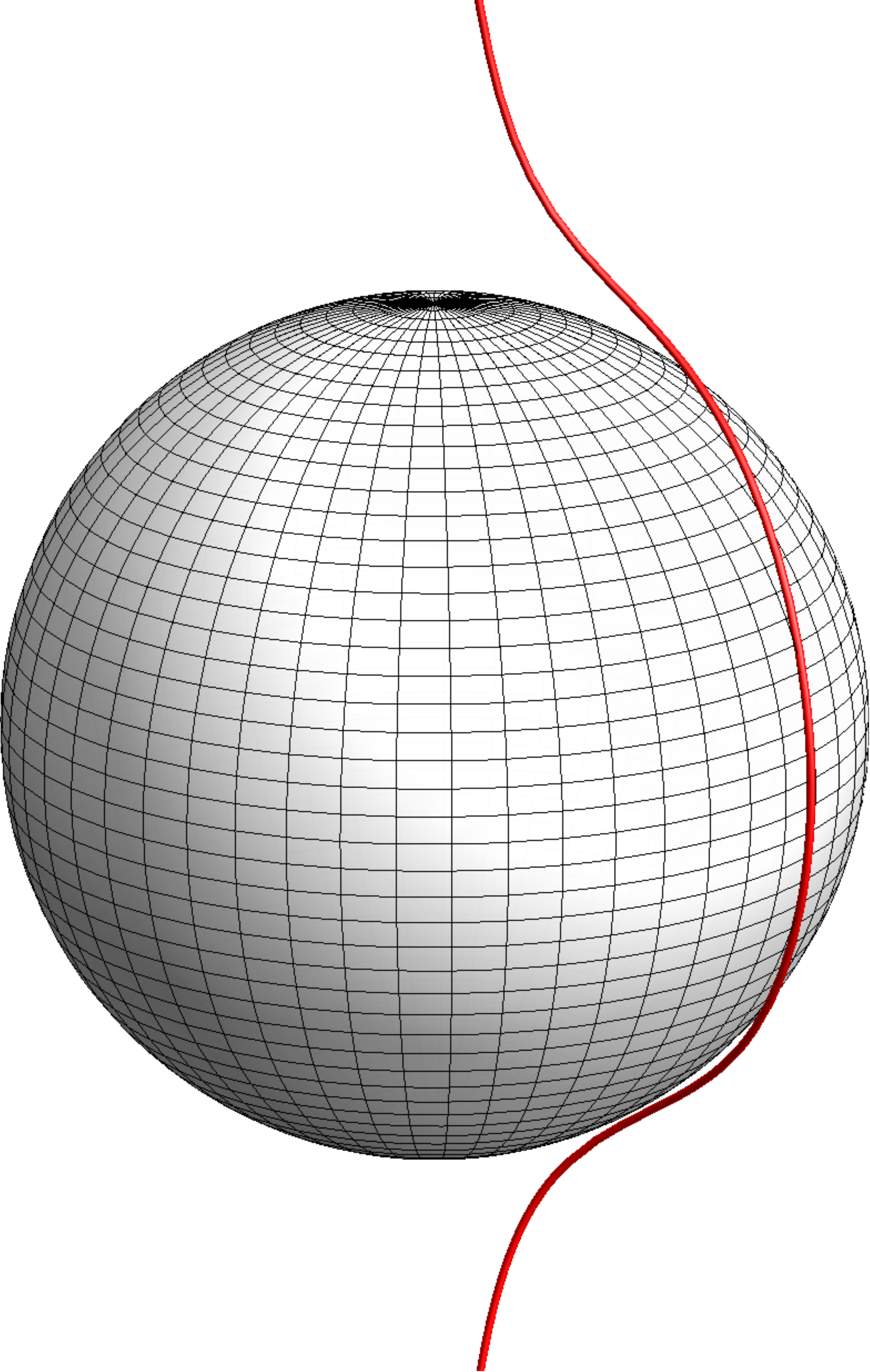}
\put(55,1){(b)}
\end{overpic}
  \caption{(a): A couple of arbitrarily chosen magnetic field lines (red) for the Wald solution perturbed by a $\Phi_0$ given by Eq.~\eqref{eq:Toy} ($l=10$). The gray sphere represents the black hole horizon. The pre-loops can become properly closed (if not already so) through magnetic reconnection later on, within the jet. Note, although these individual magnetic field lines are not axisymmetric, there are others that are simple shifts of them in the azimuthal direction, so the magnetic field as a whole is axisymmetric. (b): A similar field line for the unperturbed Wald solution. The field line does not penetrate the horizon and does not form a pre-loop. 
  \vspace{5mm}
}
	\label{fig:BFieldPert}
\end{figure}

One is then faced with the question of what mechanism is responsible for launching these magnetic loops, and for creating the disordered environment they reside in. 
A particularly interesting observation is that the randomness in the magnetic field \emph{decreases} as we move outwards along the jet (\cite{2013EPJWC..6106001W}), from an $\alpha$ of zero to a few percent near the base of the jets to tens of percent at a few tens of beam widths away (\cite{2003ApJ...589..733P,2005AJ....130.1389L,2008ApJ...681L..69G}), a trend that appears inconsistent with jet flow instability induced randomness. 
ITV turbulence provides an interesting alternative, with the benefit of being already present in the central engine. 
Aside from the turbulence in the electromagnetic field, because ITV turns up and down the energy extraction efficiency, it has the innate ability to produce disturbances in the jet material's radial velocity, encouraging the plasma flow to also take up traits of turbulence right from the beginning.
The joint electromagnetic and fluid turbulences 
not only provide a possible explanation for the microvariability seen in jet emissions (randomness in the magnetic field translates directly into that of the synchrotron radiation), they are also beneficial for the random occurrences of magnetic reconnection (i.e. they increase the rate of reconnection, see \cite{1999ApJ...517..700L}), which would help birth field line loops. 
ITV perturbations to the magnetic field also fragments otherwise smooth/straight magnetic field lines, bending them into pre-loops to be swept down the jet, so that the reconnection results are more likely to be loops rather than redirected open lines. Fig.~\ref{fig:BFieldPert} depicts a couple of arbitrary magnetic field lines of a Wald solution perturbed by a toy $\Phi_0$ as given by Eq.~\eqref{eq:Toy} (with $l=10$). As more loops are formed and subsequently sheared further down stream, the magnetic field becomes more ordered into the poloidal direction, producing the aforementioned trend of decreasing randomness. Furthermore, this process occurs where energy extraction is the most efficient, 
so it is not unreasonable that the resulting loops are abundant in the more energetic spine of the jets. In contrast, if loops are produced by kinetic instabilities for example, they would congregate near jet boundaries. 

Finally, notice that in Fig.~\ref{fig:FluxUnderPert} (a), the perturbed field lines are not symmetric against the equatorial plane, indicating that ITV would likely feed energy into the two jets asymmetrically, even causing jets to become intrinsically one-sided given the right conditions. 
Observationally, one-sided or highly asymmetric jets are not rare (see \cite{1984RvMP...56..255B} and in particular \cite{2001ApSSS.276..255S,2007MNRAS.376.1341R} for Cygnus X-1). Although relativistic beaming may help explain the faintness of a mirror jet in some cases, some level of intrinsic asymmetry can not be ruled out (see Sec.~II~A~4 of \cite{1984RvMP...56..255B}). For example, interleaving regions of emission in jets on alternative sides have been seen in a number of sources, suggesting a flip-flopping of an intrinsically one-sided jet (\cite{1984ApJ...279...74R}). 

\subsection{Manifestation in disk emission} \label{sec:PSD}
\begin{figure}[t,b]
\begin{overpic}[width=0.95\columnwidth]  {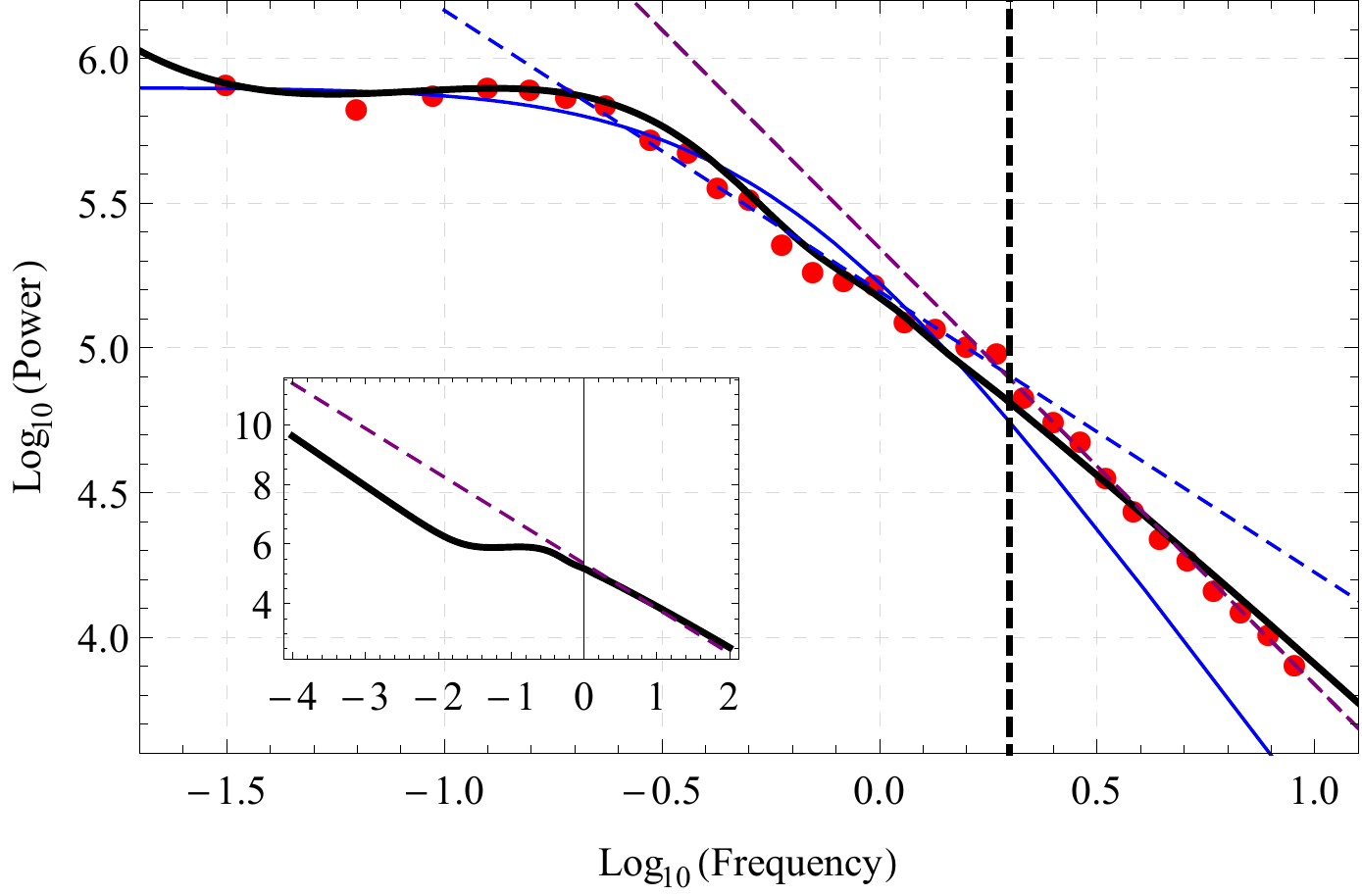}
\end{overpic}
  \caption{The red dots are the microvariation PSD data for Cygnus X-1 presented in Fig.~2 of \cite{Uttley:2001qp}. The black dashed vertical curve marks out the boundary between two ``phases''. The blue and purple dashed lines are local linear fits to segments of data close to the boundary, with slopes $\approx -1$ and $\approx -1.5$ respectively. The thin blue curve is a fit using shot noise model with a power law decay function of power $p=6.26$. The thick black curve is a fit with the new hot band model having a leakage profile \eqref{eq:LeakageTime} and a red underlying bombardment spectral density. The inset is a zoom-out of the thick black curve (and the purple dashed line for reference), showing that it returns to a power law at very low frequencies. In this figure, the frequency is in units of $\rm Hz$, and the power in units of $\rm counts^2 s^{-2} Hz^{-1}$. 
}
	\label{fig:FlatteningData}
\end{figure}

Let's now turn to examine the detailed shape of the microvariation power spectral density (PSD) associated with X-ray observations. We will formulate our arguments within the context of disk/corona-based inverse-Compton scattering (\cite{1990A&ARv...2..125B}), but note that jet emission may also contribute (\cite{2004Sci...306..998G}), and our discussion can be adapted to that case as well. For concreteness, we take the very clean data obtained for Cygnus X-1 and presented in Fig.~2 of \cite{Uttley:2001qp}, which we partially (we only take their first flux quartile as the fourth quartile curve is similar, and we do not need the error bars that are very small) reproduce in Fig.~\ref{fig:FlatteningData}. The PSD exhibits a rich structure.  
For example, there appears to be a break in the data at around $\omega=2$Hz marked out with the black dashed vertical line in Fig.~\ref{fig:FlatteningData}. The points to the right has a steeper linear fit than the points immediately to the left, with slopes of $s_r=-1.51$ and $s_l=-0.97$, respectively. And more importantly, we will have to find a process to explain the PSD's white-turning-into-red shape (\cite{1995xrbi.nasa..252V}) to the left of the dashed line.

\begin{figure}[t,b]
\centering
\begin{overpic}[width=0.9\columnwidth]  {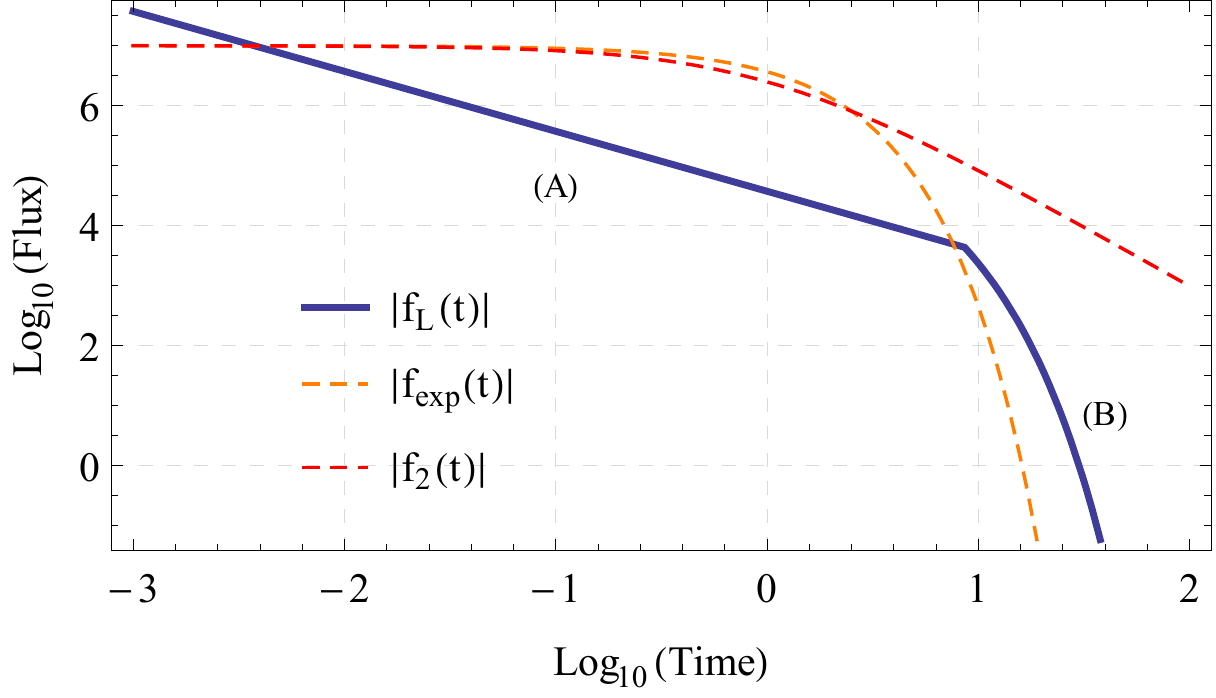}
\put(95,2){(a)}
\end{overpic}
\begin{overpic}[width=0.9\columnwidth]  {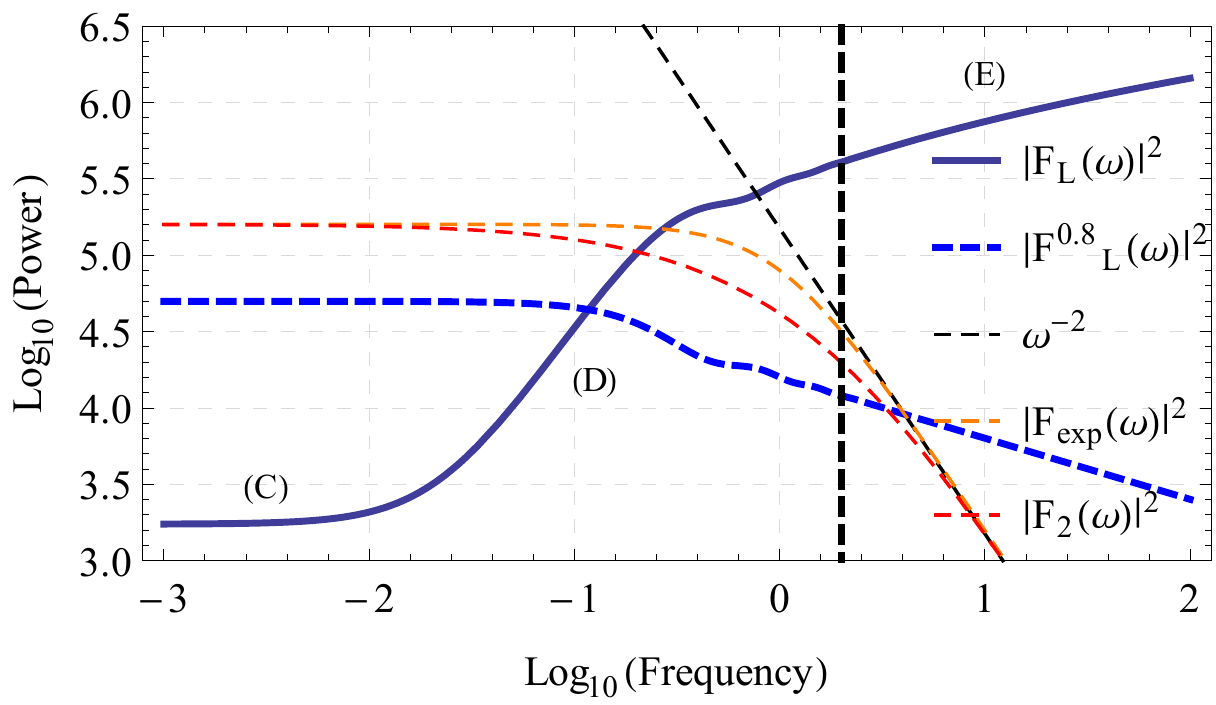}
\put(95,2){(b)}
\end{overpic}
  \caption{(a) The leakage/decay profile $f(t)$ in the time domain (c.f. Fig.~1 in \cite{1997ApJ...480..735K}) for the leakage flux $f_{\rm L}(t)$ and the decays $f_{\rm exp}(t)$ (used in the shot noise model of \cite{1972ApJ...174L..35T}) and power law $f_{2}(t)$. For $f_{\rm L}(t)$, the segments \texttt{(A)} and \texttt{(B)} correspond to the top and bottom lines in Eq.~\eqref{eq:LeakageTime}. 
  (b) The frequency domain leakage profile $|F_{\rm L}(\omega)|^2$ and its decay counterparts $|F_{\rm exp}(\omega)|^2$ and $|F_{2}(\omega)|^2$. The vertical dashed black line corresponds to the one in Fig.~\ref{fig:FlatteningData}. For $|F_{\rm L}(\omega)|^2$, the three segments \texttt{(C)}, \texttt{(D)} and \texttt{(E)} correspond to the three pieces with different slopes in the inset of Fig.~\ref{fig:FlatteningData}. 
When the power index in Eq.~\eqref{eq:LeakageTime} is increased from $-1$ as in the case of $|F^{0.8}_{\rm L}(\omega)|^2$, for which the index is changed to $-0.8$, the profile becomes similar to the decays. For the decays, the exponential decay profile 
$|F_{\rm exp}(\omega)|^2$ has a flat plateau at low frequencies, and a fixed slope of $-2$ at high frequencies (see the the black dashed fitting line). Power law decay profiles (red dashed curve) share the same general shape, and the same slope of $-2$ at high frequencies, but can have a smoother transition region. 
For both figures, the vertical axes are shifted arbitrarily so all curves fit into the same plot. 
  }
	\label{fig:FlatteningLeak}
\end{figure}

A leading explanation is the shot noise model of \cite{1972ApJ...174L..35T}, which we will describe in a little detail as we will be using essentially the same mathematics. In this model, a white noise background PSD is produced by uncorrelated shots occurring within the disk or corona, which then decay away over time. One possible physical realization of the shots is spontaneously arising flares (e.g. the magnetic flares of \cite{1999MNRAS.306L..31P}) in the corona that dies away gradually. The effect of the decay is that it is a slower process (as compared to the instantaneous leading edge of the shots) that smears out the higher frequency components of the shots, resulting in a red/pink PSD for the emitted X-ray radiation at higher frequencies, while leaving the lower frequencies untouched and white. Mathematically, let the decay profile of a flare occurring at $t=0$ be modelled by some $f(t)$, then its effect on the PSD is accounted for by a temporal convolution of $f(t)$ into the original flaring time series (a train of delta functions), which translates into a frequency space multiplication by $F(\omega)$, the Fourier transform of $f(t)$. So the overall PSD is multiplied by $|F(\omega)|^2$. 
In \cite{1972ApJ...174L..35T}, a decay is ``arbitrarily" chosen to be of an exponential form, $f_{\rm exp} = \exp(-\lambda t)\Theta(t)$, where $\Theta$ is the Heaviside step function. We see that its Fourier transform $F_{\rm exp}=1/[\sqrt{2 \pi } (\lambda -i \omega )]$ naturally provides the desired flat plateau in $|F_{\rm exp}|^2$ at lower frequencies (see Fig.~\ref{fig:FlatteningLeak}(b)), and so the choice is not entirely arbitrary. Nevertheless, we note that other decay profiles, with for example $f_p(t)=(t+1)^{-p}\Theta(t)$, have the same general $|F|^2$ shape in frequency domain, but possibly more smooth transition regions (see Fig.~\ref{fig:FlatteningLeak}(b)). With all of these decay profiles however, the slope of $|F|^2$ at high frequencies is fixed at $-2$, quite different from  the $s_l\approx -1$ and $s_r \approx -1.5$ values. This discrepancy is cited as the major shortcoming of the shot noise model. By using power law generalizations, one can trick the model to produce better fits to data, by placing the wider and smoother transition regions of $|F|^2$, instead of the steady steeply sloped region, over the observational data. The best fit (to data on the left of the dashed line only, fitting to all data leads to worse-looking results) is produced with $p=6.26$ (the solid thin blue curve in Fig.~\ref{fig:FlatteningData}). With this generalization, we achieve a decent fit for the data points at low frequencies, but the slope is unsurprisingly too high at higher frequencies. 

Alternative generalizations introducing a wide range of 
different decay time-scales (\cite{1989ESASP.296..499L,1991ApJ...376..295L}) can possibly achieve much better fits, as one would expect for an increased number of available tunable parameters, but they would also have a flat region extending all the way into extremely low frequencies. In contrast, it has been observed that (at least for galactic X-ray sources) a power-law PSD resumes at frequencies much lower than the critical value at which transition from white to red noises occurs in Fig.~\ref{fig:FlatteningData}, as possibly hinted at by the up-tick in the left-most data point in Fig.~\ref{fig:FlatteningData} 
(the extent of these data into low $\omega$'s is limited by observation-duration constraints), so the flat region is only a shoulder in the broader spectrum (\cite{1997ApJ...480..735K}). 
In addition, the standard shot noise model (and some other models such as self-organized criticality) is also incompatible with the flux dependence of the variation amplitude discussed in Sec.~\ref{sec:VarFluxCorr} (\cite{2004ApJ...612L..21G}). 
We now introduce an alternative proposal that achieves a tighter fit (with only one additional physically motivated adjustable parameter) and a return to a power law at very low frequencies. 

Specifically, energy extracted from the black hole can travel along magnetic field lines in the form of Alfv\'en waves 
\footnote{The waves propagating inside a force-free magnetosphere abutting the thin vacuum layer can be divided into two categories, the fast-magnetosonic waves and the Alfv\'en waves. A feature worth noting is that the 
Alfv\'en waves are good candidates for carrying energy out of the central engine region (see \cite{Brennan:2013jla,Zhang:2015aga} for an exact analytical solution that does so efficiently), as their propagation is regulated to be along magnetic field lines, rather than scatter randomly. Fast-magnetosonic waves on the other hand propagate more like vacuum electromagnetic waves.} 
(for rapidly-spinning black holes, the disk can get very close to the horizon, and so penetrate inside the magnetosphere, where plasma waves are allowed to propagate). Since the field lines thread through the accretion disk, energy flows would bombard the disk and corona, heating them up to create hot bands. The turbulent ITV would then cause not only a changing energy extraction rate at the source, but also a focusing and de-focusing of Alfv\'en waves by altering how tightly field lines bundle. The consequence is that the bombardment takes on a stochastic appearance, with a PSD consistent with the electromagnetic turbulence discussed previously. For this section, we take it to be a Kolmogorov $\omega^{-5/3}$ power law (but noting that making this more flexible could result in even better fits to data). 
Note that the average high frequency slopes of $s_r\approx -1.5$ for Cygnux X-1 and $\approx -1.55$ in similar PSDs for AGNs measured by \cite{1993ApJ...414L..85L} are inconsistent with either the standard shot-noise model ($-2$) or the traditional flicker noise ($-1$), but close to $-5/3$. 

The hot bands can then serve as sources for soft photons and hot electrons, replacing the accretion shocks in the inhomogeneous hot cloud model of \cite{1997ApJ...480..735K}. The hot cloud model is constructed to explain the observed time lags between hard and soft photons, as well as the power spectrum shape. Specifically, \cite{1997ApJ...480..735K} propose a Comptonization-based emission mechanism that requires the existence of large hot clouds with most of the energy at the outer boundary (in order to match to the frequency dependence of the lags). This model has been criticized on the grounds that such a peculiar energy distribution is difficult to produce physically, if one assumes that the energy sponsor for the clouds is internal to the disk. However, with our external energy source, it is natural for hot bands/clouds to be more energetic on their boundaries that are more exposed to bombardments. Since it takes time for X-ray photons to escape the hot cloud (in particular, hard photons need to experience more scattering inside the cloud to pick up energy and thus emerge later), a leakage profile needs to be overladen onto the energy bombardment profile to produce the final radiation PSD, and the mathematical details become similar to the shot noise model (convolution in the time domain translating into multiplications in the frequency domain), but the underlying white shot noise curve (also assumed for the original hot cloud model, see below) is replaced by a red $\omega^{-5/3}$ power law, and a leakage profile replaces the decay profile. 

For the photon leakage, simulations in \cite{1997ApJ...480..735K} (see Fig.~1 therein) produced results mostly hugging the $f_{\rm L}(t)=t^{-1}\Theta(t)$ profile at small $t$. 
At larger $t$, the simulations exhibit a transition from the power law into an exponential cutoff that declines much faster. 
We denote the time when this occurs as $C_1$, so the leakage profile is 
\bea \label{eq:LeakageTime}
f_{\rm L}(t) = 
\begin{cases} 
A t^{-1}\,, & 0 \leq t < C_1 \,,
\vspace{1mm}
\\ 
A t^{-1}e^{-\lambda(t-C_1)}\,, &  t \geq  C_1 \,.
\end{cases}
\eea
In the frequency domain, we then have
\begin{align}\label{eq:ToFit}
F_{\rm L}&(\omega)=
A' \bigg\{
 \Big[
\log (B \omega)+\gamma 
-\text{Ci}\left(B \omega  C_1'\right)-i \text{Si}\left(B \omega  C_1'\right)\Big] 
\notag \\ & 
+e^{C_1' \lambda '}  \Big[\text{Ei}\left(\left(i B \omega -\lambda '\right) C_1'\right)
-\text{Ei}\left(\left(i B \omega -\lambda '\right) C_2'\right)\Big]\bigg\}\,,
\end{align}
where $\gamma$ is the Euler's constant, while Si, Ci and Ei are the sine, cosine and exponential integral functions. Note we have introduced a $B$ factor multiplying onto $\omega$, which makes it easier to move $F_{\rm L}(\omega)$ across frequencies by hand to zero in on an initial guess for the fitting. This corresponds to a rescaling of $t$ by multiplying it with $1/B$, resulting in the relationships $\lambda = \lambda'/B$ and $C_1 = B C_1'$. We have also introduced an overall cutoff at $C'_2 \gg C'_1$ (i.e. $f(t)=0$ exactly when $t>BC_2'$), which helps with the Fourier transformation routine, but has no physical effects as the exponential suppression factor has already ensured that $f_{\rm L}(t) \approx 0$ at $t= BC'_2$, so $C_2'$ can be chosen to have any arbitrary large value. Adjusting the remaining parameters in Eq.~\eqref{eq:ToFit} via a simple selective (accepting a step only when it leads to an improvement) random walk procedure produces the best-fit PSD as the thick black curve in Fig.~\ref{fig:FlatteningData}, corresponding to the parameter values (leaving out the unimportant overall amplitude $A'$, which is degenerate with the amplitude of the underlying bombardment profile)
\bea
B \approx 15.0\,, \quad \quad \lambda'\approx 4.98\,,\quad \quad C'_1 \approx 0.577\,,
\eea
translating into 
\bea \label{eq:FittingVals}
\lambda \approx 0.333\,,\quad \quad 
C_1 \approx 8.63s \,.
\eea 
The $f_{\rm L}(t)$ and corresponding $F_{\rm L}(\omega)$ profiles with these parameters are shown in Fig.~\ref{fig:FlatteningLeak}. We note in particular that the $C_1$ value in Eq.~\eqref{eq:FittingVals} is consistent with the results of \cite{1997ApJ...480..735K}, obtained using realistic astrophysical parameters for Cygnus X-1. 

We reiterate that although our leakage profile is similar to those from the simulations of \cite{1997ApJ...480..735K}, the power to $t$ in our $f_{\rm L}(t)$ is precisely $-1$, which creates a more interesting profile for $F_{\rm L}(\omega)$ as shown in Fig.~\ref{fig:FlatteningLeak}(b). In the log-log plots	, the profiles of $|F_{\rm L}(\omega)|^2$ and the bombardment profile of $\omega^{-5/3}$ (a straight line with slope $-5/3$) combine by addition, so the three segments marked \texttt{(C)}, \texttt{(D)} and \texttt{(E)} in Fig.~\ref{fig:FlatteningLeak}(b) create three regimes of different slopes (see the inset in Fig.~\ref{fig:FlatteningData}). Piece \texttt{(D)} neutralizes the bombardment slope $-5/3$, creating the flat shoulder, while segments \texttt{(C)} and \texttt{(E)}, being much flatter themselves, have less impact. 
In contrast, \cite{1997ApJ...480..735K} had leakage profiles whose power indices are close to, but a little more positive than $-1$, resulting in simpler $|F_{\rm L}(\omega)|^2$'s (see the dashed blue curve in Fig.~\ref{fig:FlatteningLeak}(b)) that in fact more closely resemble the decay profiles like $|F_{\rm exp}(\omega)|^2$. Therefore, those leakage profiles need to be overladen onto a white shot noise profile in order to produce a white noise plateau, just as in the shot noise model. In other words, in either the shot noise model or the original hot cloud model, a white underlying shot noise has to be present in the first place, which is responsible for the plateau, and is masked at higher frequencies by shorter time-scale phenomena related to the evolution of each individual shot. At the other end of the spectrum, there is no natural way to mask the white noise at very low frequencies, so in order to return to a power law there, one has to invoke new and separate mechanisms (such as accretion rate fluctuations due to variations in the viscosity parameter (\cite{1997MNRAS.292..679L})). Our proposal works quite differently. The bombardment profile is a red noise, and nothing is intrinsically white, so the white noise shoulder is necessarily approximate and transient, and the morphology of the entire PSD can be accounted for with a single mechanism. 

\section{Conclusion} \label{sec:Conclude}
We caution that most, if not all, of the observational phenomena discussed above admit alternative explanations, mostly involving more complicated disk and corona dynamics (e.g. the existence of many different decay scales, fractal chain of flares, self-organized criticality in disk fluctuations etc), and indeed, multiple mechanisms are likely at work. The particular proposal introduced here has some features that we think are desirable and make it a useful candidate. First of all, ITV is an integral part of the fundamental energy process and propagates at the speed of light, thus capable of producing the observed large variations that evolve on a light-crossing (and not viscous) timescale (\cite{2003A&AT...22..661G}). It is also common across different celestial objects with mass scales spanning decades. Specifically, the ITV dynamics is largely governed by the properties of the black holes, whose simplicity (as exemplified by, e.g., no-hair theorems  (\cite{Chrusciel:1994sn,Heusler:1998ua}) when the holes are in isolation) leads to simpler scaling laws (an example is the numerical computation of gravitational waveforms for binary black hole coalescences, where only one simulation is needed, and then the resulting waveform applies to different mass scales after a simple rescaling of the axes), so ITV in AGNs is not expected to qualitatively differ from that in microquasars, and quantitatively should scale proportionally according to black hole mass. 
One commonality that jumps out immediately is that although we have taken Cygnus X-1 as an example in Sec.~\ref{sec:PSD}, the PSD for AGNs share remarkably similar features (see for example, Fig.~1 of \cite{2002A&A...390..431R} for blazar 0208-512). In addition, the characteristic variation timescale indeed scales with the black hole mass (\cite{1999ApJ...514..682E,2003A&AT...22..661G}). 

Although we have initiated our ITV discussions in the context of evading the Meissner effect, and thus concentrated on rapidly-spinning black holes, we note that ITV may nevertheless be present around more moderately spinning holes as well. So long as it increases the horizon flux as compared to the otherwise stationary background (not yet known at the present time), the discussions in Sec.~\ref{sec:Obs} still apply. It is however likely that ITV is less violent when teamed with lower spins. 
\cite{2010ApJ...711...50T} argues for higher spin in radio loud sources, and on the other hand, the variabilities of the radio-loud and quiet objects tend to be more violent and quiescent respectively, especially on the shorter time scales (\cite{1990A&ARv...2..125B}), which would at first sight at least, be consistent with more violent ITV for sources powered by more rapidly spinning holes.

Finally, due to the difficulty in quantitatively solving joint evolution systems containing electromagnetic, general relativistic, fluid dynamic, and radiation microphysical ingredients, some of the discussions are limited to a vastly simplified and heuristic level. We hope that further studies in the future, especially through numerical simulations, will provide more concrete predictions to be compared with data, and perhaps reveal more intriguing and unexpected properties of electromagnetism near rapidly spinning black holes. 
In particular, the ITV turbulence needs to be verified numerically and its energy spectrum better characterized, through targeted studies with sufficient accuracy to resolve small scale features, and the ability to distinguish randomness caused by ITV and numerical errors, so overzealous evolution stabilization measures are not triggered to artificially suppress ITV. The numerical code may also need to be able to account for the important structures surrounding a black hole in an astrophysical setting, to see if energy flows reflected off of e.g., the interstellar medium or the ion-supported or dusty torus,
result in a black hole bomb-like condition (see \cite{1972Natur.238..211P}, but with the BZ process replacing superradiance and the mirror becoming possibly leaky and only partially enclosing), which would benefit the self-sustainability of ITV.  

\vspace{0.2cm}

\acknowledgements
This work is supported by the National Natural Science Foundation of China Grants 11443008 and 11503003, the Fundamental Research Funds for the Central Universities Grant 2015KJJCB06, and a Returned Overseas Chinese Scholars Foundation grant. 

\bibliographystyle{apj}

\begin{thebibliography}{}
\expandafter\ifx\csname natexlab\endcsname\relax\def\natexlab#1{#1}\fi

\bibitem[{{Antonucci}(1993)}]{1993ARA&A..31..473A}
{Antonucci}, R. 1993, \araa, 31, 473

\bibitem[{Aretakis(2012)}]{Aretakis:2012ei}
Aretakis, S. 2012, arXiv:1206.6598

\bibitem[{Aretakis(2013)}]{Aretakis:2013dpa}
---. 2013, Phys. Rev., D87, 084052

\bibitem[{{Begelman} {et~al.}(1984){Begelman}, {Blandford}, \&
  {Rees}}]{1984RvMP...56..255B}
{Begelman}, M.~C., {Blandford}, R.~D., \& {Rees}, M.~J. 1984, Reviews of Modern
  Physics, 56, 255

\bibitem[{Bicak \& Dvorak(1976)}]{Bicak76}
Bicak, J., \& Dvorak, L. 1976, General Relativity and Gravitation, 7, 959

\bibitem[{{Bic{\'a}k} \& {Dvor{\'a}k}(1980)}]{1980PhRvD..22.2933B}
{Bic{\'a}k}, J., \& {Dvor{\'a}k}, L. 1980, Phys. Rev. D, 22, 2933

\bibitem[{Bicak \& Janis(1985)}]{Bicak85}
Bicak, J., \& Janis, V. 1985, Mon. Not. Roy. Astr. Soc., 212, 899

\bibitem[{Bicak \& Karas(1989)}]{Bicak89}
Bicak, J., \& Karas, V. 1989, in Proc. of the 5th Marcel Grossman Meeting on
  General Relativity, ed. D.~G. {Blair} \& M.~K. {Buckingham} (World
  Scientific: Singapore), 1199

\bibitem[{Bicak {et~al.}(2006)Bicak, Karas, \& Ledvinka}]{Bicak:2006hs}
Bicak, J., Karas, V., \& Ledvinka, T. 2006, arXiv:astro-ph/0610841, [IAU
  Symp.238,139(2007)]

\bibitem[{{Bi{\v c}{\'a}k} \& {Hejda}(2015)}]{2015arXiv151001911B}
{Bi{\v c}{\'a}k}, J., \& {Hejda}, F. 2015, ArXiv e-prints, arXiv:1510.01911

\bibitem[{{Bi{\v c}{\'a}k} {et~al.}(2007){Bi{\v c}{\'a}k}, {Karas}, \&
  {Ledvinka}}]{2007IAUS..238..139B}
{Bi{\v c}{\'a}k}, J., {Karas}, V., \& {Ledvinka}, T. 2007, in IAU Symposium,
  Vol. 238, IAU Symposium, ed. V.~{Karas} \& G.~{Matt}, 139--144

\bibitem[{{Blandford} \& {K{\"o}nigl}(1979)}]{1979ApJ...232...34B}
{Blandford}, R.~D., \& {K{\"o}nigl}, A. 1979, apj, 232, 34

\bibitem[{Blandford \& Payne(1982)}]{Blandford:1982di}
Blandford, R.~D., \& Payne, D.~G. 1982, Mon. Not. Roy. Astron. Soc., 199, 883

\bibitem[{{Blandford} \& {Znajek}(1977)}]{1977MNRAS.179..433B}
{Blandford}, R.~D., \& {Znajek}, R.~L. 1977, Mon. Not. Roy. Astron. Soc., 179,
  433

\bibitem[{{Brandenburg} {et~al.}(2015){Brandenburg}, {Kahniashvili}, \&
  {Tevzadze}}]{2015PhRvL.114g5001B}
{Brandenburg}, A., {Kahniashvili}, T., \& {Tevzadze}, A.~G. 2015, Physical
  Review Letters, 114, 075001

\bibitem[{{Bregman}(1990)}]{1990A&ARv...2..125B}
{Bregman}, J.~N. 1990, Astron. Astrophys. Rev., 2, 125

\bibitem[{Brennan {et~al.}(2013)Brennan, Gralla, \& Jacobson}]{Brennan:2013jla}
Brennan, T.~D., Gralla, S.~E., \& Jacobson, T. 2013, Class.Quant.Grav., 30,
  195012

\bibitem[{{Chamblin} {et~al.}(1998){Chamblin}, {Emparan}, \&
  {Gibbons}}]{1998PhRvD..58h4009C}
{Chamblin}, A., {Emparan}, R., \& {Gibbons}, G.~W. 1998, \prd, 58, 084009

\bibitem[{{Cho}(2005)}]{2005ApJ...621..324C}
{Cho}, J. 2005, \apj, 621, 324

\bibitem[{Chrusciel(1994)}]{Chrusciel:1994sn}
Chrusciel, P.~T. 1994, Contemp.Math., 170, 23

\bibitem[{{Contopoulos} {et~al.}(2013){Contopoulos}, {Kazanas}, \&
  {Papadopoulos}}]{2013ApJ...765..113C}
{Contopoulos}, I., {Kazanas}, D., \& {Papadopoulos}, D.~B. 2013, \apj, 765, 113

\bibitem[{{Degrange} {et~al.}(2008){Degrange}, {Superina}, {Giebels}, \&
  {Volpe}}]{2008bves.confE..16D}
{Degrange}, B., {Superina}, G., {Giebels}, B., \& {Volpe}, F. 2008, in Blazar
  Variability across the Electromagnetic Spectrum, 16

\bibitem[{{Dhalla} {et~al.}(2010){Dhalla}, {Webb}, {Bhatta}, \&
  {Pollock}}]{2010JSARA...4....7D}
{Dhalla}, S.~M., {Webb}, J.~R., {Bhatta}, G., \& {Pollock}, J.~T. 2010, Journal
  of the Southeastern Association for Research in Astronomy, 4, 7

\bibitem[{{Edelson} {et~al.}(2013){Edelson}, {Mushotzky}, {Vaughan}, {Scargle},
  {Gandhi}, {Malkan}, \& {Baumgartner}}]{2013ApJ...766...16E}
{Edelson}, R., {Mushotzky}, R., {Vaughan}, S., {et~al.} 2013, \apj, 766, 16

\bibitem[{{Edelson} \& {Nandra}(1999)}]{1999ApJ...514..682E}
{Edelson}, R., \& {Nandra}, K. 1999, \apj, 514, 682

\bibitem[{{Elvis} {et~al.}(2002){Elvis}, {Risaliti}, \&
  {Zamorani}}]{2002ApJ...565L..75E}
{Elvis}, M., {Risaliti}, G., \& {Zamorani}, G. 2002, Astrophys.\ J.\ Lett.,
  565, L75

\bibitem[{Gammie {et~al.}(2004)Gammie, Shapiro, \& McKinney}]{GammieEtAl:2004}
Gammie, C.~F., Shapiro, S.~L., \& McKinney, J.~C. 2004, Astrophys.\ J., 602,
  312

\bibitem[{{Gaskell}(2004)}]{2004ApJ...612L..21G}
{Gaskell}, C.~M. 2004, \apjl, 612, L21

\bibitem[{{Gaskell} \& {Klimek}(2003)}]{2003A&AT...22..661G}
{Gaskell}, C.~M., \& {Klimek}, E.~S. 2003, Astronomical and Astrophysical
  Transactions, 22, 661

\bibitem[{{Gaur} {et~al.}(2010){Gaur}, {Gupta}, {Lachowicz}, \&
  {Wiita}}]{2010ApJ...718..279G}
{Gaur}, H., {Gupta}, A.~C., {Lachowicz}, P., \& {Wiita}, P.~J. 2010, \apj, 718,
  279

\bibitem[{{Gibbons} {et~al.}(2013){Gibbons}, {Mujtaba}, \&
  {Pope}}]{2013CQGra..30l5008G}
{Gibbons}, G.~W., {Mujtaba}, A.~H., \& {Pope}, C.~N. 2013, Classical and
  Quantum Gravity, 30, 125008

\bibitem[{{Giebels} \& {Degrange}(2009)}]{2009A&A...503..797G}
{Giebels}, B., \& {Degrange}, B. 2009, \aap, 503, 797

\bibitem[{Goldreich \& Julian(1969)}]{Goldreich:1969sb}
Goldreich, P., \& Julian, W.~H. 1969, Astrophys.J., 157, 869

\bibitem[{{Goldreich} \& {Sridhar}(1995)}]{1995ApJ...438..763G}
{Goldreich}, P., \& {Sridhar}, S. 1995, \apj, 438, 763

\bibitem[{{G{\'o}mez} {et~al.}(2008){G{\'o}mez}, {Marscher}, {Jorstad},
  {Agudo}, \& {Roca-Sogorb}}]{2008ApJ...681L..69G}
{G{\'o}mez}, J.~L., {Marscher}, A.~P., {Jorstad}, S.~G., {Agudo}, I., \&
  {Roca-Sogorb}, M. 2008, \apjl, 681, L69

\bibitem[{Gou {et~al.}(2011)Gou, McClintock, Reid, Orosz, Steiner, Narayan,
  Xiang, Remillard, Arnaud, \& Davis}]{Gou:2011nq}
Gou, L., McClintock, J.~E., Reid, M.~J., {et~al.} 2011, Astrophys.~J., 742, 85

\bibitem[{{Grandi} \& {Palumbo}(2004)}]{2004Sci...306..998G}
{Grandi}, P., \& {Palumbo}, G.~G.~C. 2004, Science, 306, 998

\bibitem[{{Gupta} {et~al.}(2008){Gupta}, {Fan}, {Bai}, \&
  {Wagner}}]{2008AJ....135.1384G}
{Gupta}, A.~C., {Fan}, J.~H., {Bai}, J.~M., \& {Wagner}, S.~J. 2008, Astron.\
  J., 135, 1384

\bibitem[{{Hardee}(2008)}]{2008JPhCS.131a2052H}
{Hardee}, P. 2008, Journal of Physics Conference Series, 131, 012052

\bibitem[{{Hejda} \& {Bicak}(2014)}]{Bicak2015}
{Hejda}, F., \& {Bicak}, J. 2014, in WDS'14 Proceedings of Contributed Papers
  -- Physics, ed. J.~{\v{S}afr\'ankov\'a} \& J.~{Pavl\r{u}}, 48--55

\bibitem[{Heusler(1998)}]{Heusler:1998ua}
Heusler, M. 1998, Living Rev.Rel., 1, 6

\bibitem[{{Hook} {et~al.}(1994){Hook}, {McMahon}, {Boyle}, \&
  {Irwin}}]{1994MNRAS.268..305H}
{Hook}, I.~M., {McMahon}, R.~G., {Boyle}, B.~J., \& {Irwin}, M.~J. 1994, mnras,
  268, 305

\bibitem[{{Hughes}(2005)}]{2005ApJ...621..635H}
{Hughes}, P.~A. 2005, \apj, 621, 635

\bibitem[{{Hughes} {et~al.}(2011){Hughes}, {Aller}, \&
  {Aller}}]{2011ApJ...735...81H}
{Hughes}, P.~A., {Aller}, M.~F., \& {Aller}, H.~D. 2011, \apj, 735, 81

\bibitem[{Jacobson(2011)}]{Jacobson:2011ua}
Jacobson, T. 2011, Class. Quant. Grav., 28, 187001

\bibitem[{{Jones}(1988)}]{1988ApJ...332..678J}
{Jones}, T.~W. 1988, \apj, 332, 678

\bibitem[{{Junor} {et~al.}(1999){Junor}, {Biretta}, \&
  {Livio}}]{1999Natur.401..891J}
{Junor}, W., {Biretta}, J.~A., \& {Livio}, M. 1999, Nature, 401, 891

\bibitem[{Karas \& Budinova(2000)}]{Karas00}
Karas, V., \& Budinova. 2000, Phys. Scr., 61, 253

\bibitem[{{Karas} \& {Vokrouhlick{\'y}}(1991)}]{1991JMP....32..714K}
{Karas}, V., \& {Vokrouhlick{\'y}}, D. 1991, Journal of Mathematical Physics,
  32, 714

\bibitem[{{Kazanas} {et~al.}(1997){Kazanas}, {Hua}, \&
  {Titarchuk}}]{1997ApJ...480..735K}
{Kazanas}, D., {Hua}, X.-M., \& {Titarchuk}, L. 1997, \apj, 480, 735

\bibitem[{King(1977)}]{King77}
King, A.~R. 1977, Mathematical Proceedings of the Cambridge Philosophical
  Society, 81, 149

\bibitem[{King {et~al.}(1975)King, Lasota, \& Kundt}]{King1975}
King, A.~R., Lasota, J.~P., \& Kundt, W. 1975, Phys. Rev. D, 12, 3037

\bibitem[{Kinnersley(1969)}]{Kinnersley:1969zza}
Kinnersley, W. 1969, J.Math.Phys., 10, 1195

\bibitem[{{Lawrence} \& {Papadakis}(1993)}]{1993ApJ...414L..85L}
{Lawrence}, A., \& {Papadakis}, I. 1993, \apjl, 414, L85

\bibitem[{{Lazarian} \& {Vishniac}(1999)}]{1999ApJ...517..700L}
{Lazarian}, A., \& {Vishniac}, E.~T. 1999, \apj, 517, 700

\bibitem[{{Lehto}(1989)}]{1989ESASP.296..499L}
{Lehto}, H.~J. 1989, in ESA Special Publication, Vol. 296, Two Topics in X-Ray
  Astronomy, Volume 1: X Ray Binaries. Volume 2: AGN and the X Ray Background,
  ed. J.~{Hunt} \& B.~{Battrick}, 499--503

\bibitem[{{Lister} \& {Homan}(2005)}]{2005AJ....130.1389L}
{Lister}, M.~L., \& {Homan}, D.~C. 2005, \aj, 130, 1389

\bibitem[{{Lochner} {et~al.}(1991){Lochner}, {Swank}, \&
  {Szymkowiak}}]{1991ApJ...376..295L}
{Lochner}, J.~C., {Swank}, J.~H., \& {Szymkowiak}, A.~E. 1991, \apj, 376, 295

\bibitem[{Lucietti \& Reall(2012)}]{Lucietti:2012sf}
Lucietti, J., \& Reall, H.~S. 2012, Phys. Rev., D86, 104030

\bibitem[{{Lyubarskii}(1997)}]{1997MNRAS.292..679L}
{Lyubarskii}, Y.~E. 1997, \mnras, 292, 679

\bibitem[{{Lyutyj} \& {Oknyanskij}(1987)}]{1987AZh....64..465L}
{Lyutyj}, V.~M., \& {Oknyanskij}, V.~L. 1987, \azh, 64, 465

\bibitem[{McClintock {et~al.}(2013)McClintock, Narayan, \&
  Steiner}]{McClintock:2013vwa}
McClintock, J.~E., Narayan, R., \& Steiner, J.~F. 2013, arXiv:1303.1583

\bibitem[{McClintock {et~al.}(2006)McClintock, Shafee, Narayan, Remillard,
  Davis, {et~al.}}]{McClintock:2006xd}
McClintock, J.~E., Shafee, R., Narayan, R., {et~al.} 2006, Astrophys.J., 652,
  518

\bibitem[{{McKinney} \& {Gammie}(2004)}]{2004ApJ...611..977M}
{McKinney}, J.~C., \& {Gammie}, C.~F. 2004, \apj, 611, 977

\bibitem[{{McKinney} {et~al.}(2012){McKinney}, {Tchekhovskoy}, \&
  {Blandford}}]{2012MNRAS.423.3083M}
{McKinney}, J.~C., {Tchekhovskoy}, A., \& {Blandford}, R.~D. 2012, \mnras, 423,
  3083

\bibitem[{{Miller} {et~al.}(2002){Miller}, {Fabian}, {Wijnands}, {Reynolds},
  {Ehle}, {Freyberg}, {van der Klis}, {Lewin}, {Sanchez-Fernandez}, \&
  {Castro-Tirado}}]{2002ApJ...570L..69M}
{Miller}, J.~M., {Fabian}, A.~C., {Wijnands}, R., {et~al.} 2002, \apjl, 570,
  L69

\bibitem[{{Miniutti} {et~al.}(2004){Miniutti}, {Fabian}, \&
  {Miller}}]{2004MNRAS.351..466M}
{Miniutti}, G., {Fabian}, A.~C., \& {Miller}, J.~M. 2004, \mnras, 351, 466

\bibitem[{{Mocanu} \& {S{\'a}ndor}(2012)}]{2012Ap&SS.342..147M}
{Mocanu}, G.~R., \& {S{\'a}ndor}, B. 2012, \apss, 342, 147

\bibitem[{{Narayan} \& {McClintock}(2012)}]{2012MNRAS.419L..69N}
{Narayan}, R., \& {McClintock}, J.~E. 2012, \mnras, 419, L69

\bibitem[{Pan \& Yu(2015)}]{PanYu2015}
Pan, Z., \& Yu, C. 2015, arXiv:1511.07925

\bibitem[{Penna(2014)}]{Penna:2014aza}
Penna, R.~F. 2014, Phys. Rev., D89, 104057

\bibitem[{{Penna} {et~al.}(2013){Penna}, {Narayan}, \& {S{\c
  a}dowski}}]{2013MNRAS.436.3741P}
{Penna}, R.~F., {Narayan}, R., \& {S{\c a}dowski}, A. 2013, \mnras, 436, 3741

\bibitem[{{Pollack} {et~al.}(2003){Pollack}, {Taylor}, \&
  {Zavala}}]{2003ApJ...589..733P}
{Pollack}, L.~K., {Taylor}, G.~B., \& {Zavala}, R.~T. 2003, \apj, 589, 733

\bibitem[{{Poutanen} \& {Fabian}(1999)}]{1999MNRAS.306L..31P}
{Poutanen}, J., \& {Fabian}, A.~C. 1999, \mnras, 306, L31

\bibitem[{{Press} \& {Teukolsky}(1972)}]{1972Natur.238..211P}
{Press}, W.~H., \& {Teukolsky}, S.~A. 1972, \nat, 238, 211

\bibitem[{{Rees} {et~al.}(1982){Rees}, {Begelman}, {Blandford}, \&
  {Phinney}}]{1982Natur.295...17R}
{Rees}, M.~J., {Begelman}, M.~C., {Blandford}, R.~D., \& {Phinney}, E.~S. 1982,
  \nat, 295, 17

\bibitem[{Reynolds(2013)}]{Reynolds:2013rva}
Reynolds, C.~S. 2013, Class. Quant. Grav., 30, 244004

\bibitem[{{Romero} {et~al.}(2002){Romero}, {Cellone}, {Combi}, \&
  {Andruchow}}]{2002A&A...390..431R}
{Romero}, G.~E., {Cellone}, S.~A., {Combi}, J.~A., \& {Andruchow}, I. 2002,
  \aap, 390, 431

\bibitem[{{Rosen} {et~al.}(1999){Rosen}, {Hardee}, {Clarke}, \&
  {Johnson}}]{1999ApJ...510..136R}
{Rosen}, A., {Hardee}, P.~E., {Clarke}, D.~A., \& {Johnson}, A. 1999, \apj,
  510, 136

\bibitem[{{Rudnick} \& {Edgar}(1984)}]{1984ApJ...279...74R}
{Rudnick}, L., \& {Edgar}, B.~K. 1984, \apj, 279, 74

\bibitem[{{Russell} {et~al.}(2007){Russell}, {Fender}, {Gallo}, \&
  {Kaiser}}]{2007MNRAS.376.1341R}
{Russell}, D.~M., {Fender}, R.~P., {Gallo}, E., \& {Kaiser}, C.~R. 2007,
  \mnras, 376, 1341

\bibitem[{{Sasada} {et~al.}(2008){Sasada}, {Uemura}, {Arai}, {Fukazawa},
  {Kawabata}, {Ohsugi}, {Yamashita}, {Isogai}, {Sato}, \&
  {Kino}}]{2008PASJ...60L..37S}
{Sasada}, M., {Uemura}, M., {Arai}, A., {et~al.} 2008, pasj, 60, L37

\bibitem[{{Spencer} {et~al.}(2001){Spencer}, {de La Force}, {Stirling},
  {Garrett}, {Fender}, \& {Ogley}}]{2001ApSSS.276..255S}
{Spencer}, R., {de La Force}, C., {Stirling}, A., {et~al.} 2001, Astrophysics
  and Space Science Supplement, 276, 255

\bibitem[{{Steiner} {et~al.}(2013){Steiner}, {McClintock}, \&
  {Narayan}}]{2013ApJ...762..104S}
{Steiner}, J.~F., {McClintock}, J.~E., \& {Narayan}, R. 2013, \apj, 762, 104

\bibitem[{Superina \& Degrange(2008)}]{Superina:2008zz}
Superina, G., \& Degrange, B. 2008, PoS, BLAZARS2008, 066

\bibitem[{{Takamori} {et~al.}(2011){Takamori}, {Nakao}, {Ishihara}, {Kimura},
  \& {Yoo}}]{2011MNRAS.412.2417T}
{Takamori}, Y., {Nakao}, K.-I., {Ishihara}, H., {Kimura}, M., \& {Yoo}, C.-M.
  2011, \mnras, 412, 2417

\bibitem[{{Tchekhovskoy} {et~al.}(2012){Tchekhovskoy}, {McKinney}, \&
  {Narayan}}]{2012JPhCS.372a2040T}
{Tchekhovskoy}, A., {McKinney}, J.~C., \& {Narayan}, R. 2012, Journal of
  Physics Conference Series, 372, 012040

\bibitem[{{Tchekhovskoy} {et~al.}(2010){Tchekhovskoy}, {Narayan}, \&
  {McKinney}}]{2010ApJ...711...50T}
{Tchekhovskoy}, A., {Narayan}, R., \& {McKinney}, J.~C. 2010, \apj, 711, 50

\bibitem[{{Terrell}(1972)}]{1972ApJ...174L..35T}
{Terrell}, Jr., N.~J. 1972, \apjl, 174, L35

\bibitem[{Teukolsky(1973)}]{Teukolsky}
Teukolsky, S. 1973, Astrophys. J., 185, 635

\bibitem[{Teukolsky \& Press(1974)}]{Teukolsky:1974yv}
Teukolsky, S., \& Press, W. 1974, Astrophys.J., 193, 443

\bibitem[{Thorne(1974)}]{Thorne:1974}
Thorne, K.~S. 1974, Astrophys.\ J., 191, 507

\bibitem[{Thorne {et~al.}(1986)Thorne, Price, \&
  MacDonald}]{Thorne-Price-MacDonald}
Thorne, K.~S., Price, R.~H., \& MacDonald, D.~A. 1986, Black Holes: The
  Membrane Paradigm, 1st edn. (New Haven and London: Yale University Press)

\bibitem[{{Urry} \& {Padovani}(1995)}]{1995PASP..107..803U}
{Urry}, C.~M., \& {Padovani}, P. 1995, \pasp, 107, 803

\bibitem[{Uttley \& McHardy(2001)}]{Uttley:2001qp}
Uttley, P., \& McHardy, I.~M. 2001, Mon. Not. Roy. Astron. Soc., 323, L26

\bibitem[{Uttley {et~al.}(2005)Uttley, McHardy, \& Vaughan}]{Uttley:2005cc}
Uttley, P., McHardy, I.~M., \& Vaughan, S. 2005, Mon. Not. Roy. Astron. Soc.,
  359, 345

\bibitem[{{van der Klis}(1995)}]{1995xrbi.nasa..252V}
{van der Klis}, M. 1995, X-ray Binaries, 252

\bibitem[{{Vaughan} {et~al.}(2003){Vaughan}, {Edelson}, {Warwick}, \&
  {Uttley}}]{2003MNRAS.345.1271V}
{Vaughan}, S., {Edelson}, R., {Warwick}, R.~S., \& {Uttley}, P. 2003, \mnras,
  345, 1271

\bibitem[{Wald(1974)}]{PhysRevD.10.1680}
Wald, R.~M. 1974, Phys. Rev. D, 10, 1680

\bibitem[{{Wang} {et~al.}(2006){Wang}, {Chen}, {Ho}, \&
  {McLure}}]{2006ApJ...642L.111W}
{Wang}, J.-M., {Chen}, Y.-M., {Ho}, L.~C., \& {McLure}, R.~J. 2006, Astrophys.\
  J.\ Lett., 642, L111

\bibitem[{{Wardle}(2013)}]{2013EPJWC..6106001W}
{Wardle}, J.~F.~C. 2013, in European Physical Journal Web of Conferences,
  Vol.~61, European Physical Journal Web of Conferences, 6001

\bibitem[{{Wardle} {et~al.}(1994){Wardle}, {Cawthorne}, {Roberts}, \&
  {Brown}}]{1994ApJ...437..122W}
{Wardle}, J.~F.~C., {Cawthorne}, T.~V., {Roberts}, D.~H., \& {Brown}, L.~F.
  1994, \apj, 437, 122

\bibitem[{Wold {et~al.}(2007)Wold, Brotherton, \& Shang}]{Wold:2006eb}
Wold, M., Brotherton, M.~S., \& Shang, Z. 2007, Mon. Not. Roy. Astron. Soc.,
  375, 989

\bibitem[{{Yang} {et~al.}(2015){Yang}, {Zhang}, {Green}, \&
  {Lehner}}]{2015PhRvD..91h4007Y}
{Yang}, H., {Zhang}, F., {Green}, S.~R., \& {Lehner}, L. 2015, \prd, 91, 084007

\bibitem[{Yang {et~al.}(2013)Yang, Zimmerman, Zenginoğlu, Zhang, Berti,
  {et~al.}}]{Yang:2013uba}
Yang, H., Zimmerman, A., Zenginoğlu, A., {et~al.} 2013, Phys.Rev., D88, 044047

\bibitem[{{Zhang} {et~al.}(2012){Zhang}, {Brink}, {Szil{\'a}gyi}, \&
  {Lovelace}}]{2012PhRvD..86h4020Z}
{Zhang}, F., {Brink}, J., {Szil{\'a}gyi}, B., \& {Lovelace}, G. 2012, \prd, 86,
  084020

\bibitem[{Zhang {et~al.}(2015)Zhang, McWilliams, \& Pfeiffer}]{Zhang:2015aga}
Zhang, F., McWilliams, S.~T., \& Pfeiffer, H.~P. 2015, Phys. Rev., D92, 024049

\bibitem[{{Zrake}(2014)}]{2014ApJ...794L..26Z}
{Zrake}, J. 2014, \apjl, 794, L26

\bibitem[{Zrake \& East(2015)}]{ZrakeEast2015}
Zrake, J., \& East, W.~E. 2015, arXiv:1509.00461

\end{thebibliography}

\end{document}